\documentclass[%
reprint,
amsmath,amssymb,
aps,prb,
superscriptaddress,
]{revtex4-2}

\usepackage{booktabs}
\usepackage{makecell}
\usepackage{subfigure}
\usepackage{bm}
\usepackage{amsmath}
\usepackage{dcolumn}
\usepackage{float}
\usepackage{multirow}
\usepackage{color,xcolor}
\usepackage{natbib}
\usepackage{siunitx}

\usepackage{graphicx}

\usepackage{hyperref}
\hypersetup{hypertex=true,
	colorlinks=true,
	linkcolor=blue,
	anchorcolor=blue,
	citecolor=blue}

\begin{document}
	\preprint{APS/123-QED}
	\title{Evaluating Gilbert Damping in Magnetic Insulators from First Principles}
	\author{Liangliang Hong}
	\affiliation{Key Laboratory of Computational Physical Sciences (Ministry of Education), Institute of Computational Physical Sciences,\\ State Key  Laboratory  of  Surface  Physics, and Department of Physics, Fudan University, Shanghai 200433, China}
	\affiliation{Shanghai Qi Zhi Institute, Shanghai 200030, China}
	\author{Changsong Xu}
	\author{Hongjun Xiang}
	\email{hxiang@fudan.edu.cn}
	\affiliation{Key Laboratory of Computational Physical Sciences (Ministry of Education), Institute of Computational Physical Sciences,\\ State Key  Laboratory  of  Surface  Physics, and Department of Physics, Fudan University, Shanghai 200433, China}
	\affiliation{Shanghai Qi Zhi Institute, Shanghai 200030, China}
	\date{\today}
	
	\begin{abstract}
		Magnetic damping has a significant impact on the performance of various magnetic and spintronic devices, making it a long-standing focus of research. The strength of magnetic damping is usually quantified by the Gilbert damping constant in the Landau-Lifshitz-Gilbert equation. Here we propose a first-principles based approach to evaluate the Gilbert damping constant contributed by spin-lattice coupling in magnetic insulators. The approach involves effective Hamiltonian models and spin-lattice dynamics simulations. As a case study, we applied our method to Y$_3$Fe$_5$O$_{12}$, MnFe$_2$O$_4$ and Cr$_2$O$_3$. Their damping constants were calculated to be $0.8\times10^{-4}$, $0.2\times10^{-4}$, $2.2\times 10^{-4}$, respectively at a low temperature. The results for Y$_3$Fe$_5$O$_{12}$ and Cr$_2$O$_3$ are in good agreement with experimental measurements, while the discrepancy in MnFe$_2$O$_4$ can be attributed to the inhomogeneity and small band gap in real samples. The stronger damping observed in Cr$_2$O$_3$, compared to Y$_3$Fe$_5$O$_{12}$, essentially results from its stronger spin-lattice coupling. In addition, we confirmed a proportional relationship between damping constants and the temperature difference of subsystems, which had been reported in previous studies. These successful applications suggest that our approach serves as a promising candidate for estimating the Gilbert damping constant in magnetic insulators.
	\end{abstract}
	
	\maketitle
	
	\section{introduction}\label{sec-I}
	Recent decades have witnessed rapid developments in magnetics and spintronics \cite{WOS001, WOS002, WOS003}. A long-time pursuit in spintronics is to actively control and manipulate the spin degrees of freedom in solid-state systems. Related fundamental studies involve spin transport, spin dynamics and spin relaxation \cite{WOS004}. Within these domains, magnetic damping often plays a crucial role. Generally, stronger damping enables a faster writing rate for magnetic memories, while lower damping leads to a longer propagation distance of spin waves. Therefore, it is always essential to accurately evaluate the magnetic damping in different materials. For instance, yttrium iron garnet (YIG) is a highly promising spintronic material due to its ultra-low magnetic damping \cite{RN164, RN163, RN285}. However, the intrinsic mechanism behind its unique property has yet to be fully elucidated, which partly motivates us to carry out this work.
	
	At present, magnetic damping is typically represented by a phenomenological term in the well-known Landau-Lifshitz-Gilbert (LLG) equation, which has been widely employed to simulate magnetization dynamics \cite{landau1935, RN157}. A basic form of this equation can be written as, 
	\begin{align}\label{eq-1}
		\frac{\partial\vec{m}}{\partial t} = -\gamma \vec{m}\times\vec{B}+\frac{\alpha}{m} \vec{m}\times \frac{\partial{\vec{m}}}{\partial t}
	\end{align}
	where $\vec{B}$ represents the total magnetic field acting on the local dipole $\vec{m}$, $m$ denotes the norm of $\vec{m}$, $\gamma$ is the gyromagnetic ratio, and $\alpha$ is the Gilbert damping constant. The second term on the right side, as we mentioned, leads directly to the relaxation process, in which the rate of energy dissipation is determined by the damping constant. Given the importance of $\alpha$ in magnetization dynamics, its origin has been extensively studied in the literature \cite{RN146, RN156, RN206, RN209}. To the best of our knowledge, both intrinsic and extrinsic mechanisms contribute to the damping. Specifically, the intrinsic factors include spin-lattice and spin-electron couplings, while the extrinsic contributions primarily involve lattice imperfections and eddy currents \cite{RN138, RN345}. 
	
	Two types of first-principles based methods have been developed to calculate the damping constants in the past. One approach involves the breathing Fermi surface model \cite{RN308, RN310} and the torque correlation model \cite{RN311, RN309}, while the other is based on the scattering theory from linear response \cite{RN143, RN343, RN344}. These methods have demonstrated remarkable success in studying the magnetic damping in transition metals such as Fe, Co, and Ni. Despite being free from complicated experiments, which are mostly based on ferromagnetic resonance, these theoretical approaches still exhibit several limitations. Firstly, when dealing with complex systems, we often have to spend a significant amount of computing resources on the first-principles calculations. In addition, these methods are more suitable for calculating the electronic contribution to Gilbert damping in metallic magnets, thus rarely taking the effect of spin-lattice coupling into consideration \cite{RN138, RN267}.
	
	Recently, spin-lattice dynamics (SLD) simulations \cite{tranchida2018} have been adopted as an alternative method to evaluate the Gilbert damping parameters. In Ref.\ \cite{RN267}, the authors constructed an empirically parameterized Hamiltonian model for a cobalt cluster. They coupled a pre-heated lattice with a fully ordered spin state, then performed SLD simulation. During the relaxation process, the energy of lattice and spin subsystems were recorded and fitted to the following logistic functions, 
	\begin{align}
		U_{\text{lat}} &= U_0^{\text{lat}} - \frac{\Delta U_0}{1 + \text{exp}[-\eta\Delta U_0 t - \Theta]} \\
		U_{\text{mag}} &= U_0^{\text{mag}} + \frac{\Delta U_0}{1 + \text{exp}[-\eta\Delta U_0 t - \Theta]}
	\end{align}
	from which they extracted the relaxation rate $\Gamma = \eta \Delta U_0$ and calculated the damping constant $\alpha = \eta \mu_S /\gamma$. Here, $\mu_S$ denotes the magnitude of magnetic moments. In Ref.\ \cite{RN269}, the authors also built an empirical potential model for a periodic bcc Fe system. They firstly applied an external magnetic field in the z-direction and thermalized the system to a finite temperature. Then, the magnetization orientation of each atom was rotated artificially by a same angle. Afterwards, the system would relax back to equilibrium, during which the averaged z component of atomic magnetization was recorded and fitted to the following function,
	\begin{align}
		m_z(t) = \tanh \left[\frac{\alpha}{1+\alpha^2} \gamma B_{ext} (t + t_0) \right]
	\end{align}
	where $\alpha$ was exactly the Gilbert damping parameter to be estimated. Since these works selected transition metals as the research object, their results were both orders of magnitude smaller than the experimental values. In addition, the use of empirically parameterized models reduced the accuracy of their simulated results.

    In this work, we combine SLD simulations with first-principles based effective Hamiltonian models to evaluate the damping constants in magnetic insulators, where the dominant contribution results from spin-lattice couplings. Compared to the previous studies, our work has made improvements mainly in two aspects. Firstly, the utilization of first-principles based Hamiltonian models in simulations enhances the reliability of our conclusions. Besides, the better choice of research objects allows for demonstrating the superiority of SLD simulations. In particular, the microscopic origin of low damping in YIG will be investigated. The paper is organized as follows. In Sec.\ \ref{sec-II}, we introduce our effective Hamiltonian model, parameterization methods, and a scheme for evaluating Gilbert damping parameters. Then, both the validation and application of our method are presented in Sec.\ \ref{sec-III}. Finally, we summarize this work and give a brief outlook in Sec.\ \ref{sec-IV}.
	
	\section{model and methods}\label{sec-II}
	This section is split into three parts. Firstly (in Sec.\ \ref{sec-II-A}), we introduce a generic form of our effective Hamiltonian model. Then, methods involving the calculation of model parameters are presented in Sec.\ \ref{sec-II-B}. At the last part (Sec.\ \ref{sec-II-C}), we propose a novel scheme to determine the Gilbert damping constant through dynamics simulations.
	
	\subsection{The Hamiltonian Model}\label{sec-II-A}
	Since our purpose is to evaluate the contribution of spin-lattice coupling to magnetic damping, the effective Hamiltonian model must incorporate both spin and lattice degrees of freedom. A concise and generic formula that meets our basic requirements consists of the three terms as follows:
	\begin{align}
		H = H_L\left(\{u_{i,\alpha}\}\right) + H_S\left(\{\vec{s}_j\}\right) + H_{SLC}\left(\{u_{i,\alpha},\vec{s}_j\}\right)
	\end{align}
	where $\alpha$ abbreviates three orthogonal axes, $u_{i,\alpha}$ represents the displacement of atom $i$, and $\vec{s}_j$ is a unit vector that represents the direction of spin $j$.
	
	The first term $H_L$ in Hamiltonian model describes the dynamical behavior of individual phonons. Technically, we take the atomic displacements as independent variables and expand the Hamiltonian to the second order with Taylor series. Then, we have the form as,
	\begin{align}
		H_L = \frac{1}{2} \sum_{ij}\sum_{\alpha\beta} K_{ij,\alpha\beta} u_{i,\alpha} u_{j,\beta} + \frac{1}{2} \sum_{i,\alpha} M_i \dot{u}_{i,\alpha} \dot{u}_{i,\alpha} 
	\end{align}
	where $K_{ij,\alpha\beta}$ denotes the force constant tensor and $M_i$ represents the mass of atom $i$.
	
	Similarly, the second term $H_S$ describes the dynamical behavior of individual magnons. For simplicity but no loss of accuracy, we only considered the Heisenberg exchange interactions between neighbor magnetic atoms in this work, though more complex interactions could be taken into account in principle. Therefore, this term can be expressed as,
	\begin{align}
		H_S = \sum_{\langle i,j\rangle} J_{ij} \vec{S}_i\cdot\vec{S}_j
	\end{align}
	where $J_{ij}$ denotes the isotropic magnetic interaction coefficient.
	
    The third term $H_{SLC}$ represents the coupling of spin and lattice subsystems, and is expected to describe the scattering process between phonons and magnons. As an approximation of the lowest order, this term can be written as,
	\begin{align}
		H_{SLC} = \sum_{\langle i,j\rangle}\sum_{k\alpha} \left(\frac{\partial J_{ij}}{\partial u_{k,\alpha}} u_{k,\alpha}\right) \vec{S}_i\cdot\vec{S}_j
	\end{align}	
	
	According to the theory of quantum mechanics, this coupling term provides a fundamental description of the single-phonon scattering process, which is believed to be dominant among all scatterings in the low-temperature region. This type of relaxation mechanism in ferromagnetic resonance was systematically studied by Kasuya and LeCraw for the first time \cite{RN306}. It's worth noting that a higher order of Taylor expansion could have been conducted to improve the accuracy of Hamiltonian models directly. For instance, the scattering between individual phonons can be adequately described by the anharmonic terms. However, as one always has to make a trade-off between the precision and complexity of models, in this work we choose to neglect the high order terms since the anharmonic effects in current investigated systems are not important.
	
	In this study, we adopted the symmetry-adapted cluster expansion method implemented in the Property Analysis and Simulation Package for Materials (PASP) \cite{RN40} to build the Hamiltonian model presented above. This package can identify the nonequivalent interactions and equivalent atom clusters in a crystal system by analyzing its structural properties based on the group theory. A significant benefit of working with PASP is we are enabled to describe the target system with the least number of parameters. In the next section, we will discuss how to calculate the model parameters for different materials. 
	
	\subsection{Calculation of Model Parameters}\label{sec-II-B}
	Firstly, the Heisenberg exchange coefficients $J_{ij}$ and spin-lattice coupling constants $\partial J_{ij}/\partial u_{k,\alpha}$ can be calculated with the four-state method \cite{RN288, RN96}. The basic flow is to construct four artificially designated spin states of the target system, calculate the corresponding energies and forces based on the density functional theory (DFT), then determine the parameters by proper combination of those results. At the last step, the following formulas will be used,
	\begin{align}
		J_{ij} &= \frac{E^{\uparrow\uparrow} + E^{\downarrow\downarrow} - E^{\uparrow\downarrow} - E^{\downarrow\uparrow}}{4 S^2}\\
		\frac{\partial J_{ij}}{\partial u_{k,\alpha}} &= \frac{F^{\uparrow\uparrow}_{k,\alpha} + F^{\downarrow\downarrow}_{k,\alpha} - F^{\uparrow\downarrow}_{k,\alpha} - F^{\downarrow\uparrow}_{k,\alpha}}{4 S^2}
	\end{align}
	where $S$ is the spin quantum number of magnetic atoms, $E$ is the total energy of system and $F_{k,\alpha}$ refers to one component of the force on atom $k$. The superscripts ($\uparrow\uparrow$, $\downarrow\downarrow$, $\uparrow\downarrow$, $\downarrow\uparrow$) specify the constrained spin states of system in the calculation. More technical information about the four-state method can be found in the references \cite{RN288, RN96}. Compared to other approaches, the four-state method offers an obvious advantage in that no additional DFT calculations are needed to determine the coupling constants $\partial J_{ij}/\partial u_{k,\alpha}$ once the exchange coefficients $J_{ij}$ have been obtained. This is because the energy and forces are typically provided simultaneously by one DFT calculation.

	Since atomic masses $M_i$ can be directly obtained from the periodic table, more efforts are needed to deal with the force constant tensor $K_{ij,\alpha\beta}$. Currently, there are two commonly adopted ways to calculate the force constant tensor: density functional perturbation theory (DFPT) and finite displacement method. Both of these methods are applicable to our task. 
	
	However, we cannot directly take the force constant tensor obtained from first-principles calculations as the model parameter. This is because in dynamics simulations we usually expand crystal cells to reduce the undesired influence of thermal fluctuations, which leads to a conflict between the periodic boundary condition and the locality (also known as nearsightedness \cite{RN291,RN79}) of models. To be more specific, when calculating the contribution of one atom or spin to the total energy, we tend to set a well designed cutoff radius and ignore the interactions beyond it. This step is essential when dealing with a large-scale system, otherwise we will suffer from the model complexity and the computational cost. Nevertheless, if we set the elements of $K_{ij,\alpha\beta}$ that represent out-of-range interactions to be zero and leave the others unchanged, we may violate the so-called acoustic summation rules:
	\begin{align}\label{eq-4}
		\sum_{i} K_{ij,\alpha\beta} = 0 \quad \text{for all } j,\alpha,\beta.
	\end{align}
	
	It should be pointed out that a straightforward enforcement of the acoustic summation rules, achieved by subtracting errors uniformly from force constants, will break the inherent crystal symmetry inevitably, which is the technique employed in phonopy \cite{RN296}. To address the above issues, we adopted a more appropriate method in this work. Before a detailed introduction, it's necessary to recall that not every element of the force constant tensor serves as an independent variable due to the crystal symmetries. Taking the cubic cell of Y$_3$Fe$_5$O$_{12}$ (containing 160 atoms) for example, there are 230400 elements in the tensor. After symmetry analyses, we find that only 597 independent variables $\{p_n\}$ are needed to adequately determine all the tensor elements $\{K_{ij,\alpha\beta} \left(\{p_n\}\right) \}$, where the effect of locality is already considered. Afterwards, our method is to set a correction factor $x_n$ for each variable $p_n$ and minimize the deviation of parameters under the constraints of Eq.\ (\ref{eq-4}). A mathematical reformulation of this method can be written as,
	\begin{equation}\label{eq-5}
	\begin{aligned}
		& \min_{\{x_n\}}\ \sum_n (x_n - 1)^2, \ \text{with }\\
		& \sum_{i} K_{ij,\alpha\beta}(\{x_n p_n\}) = 0 \quad \text{for all } j,\alpha,\beta.
	\end{aligned}
	\end{equation}
	In the case of Y$_3$Fe$_5$O$_{12}$, there are only 18 linearly independent constraints, which allow the extremum problem to be solved rigorously. The modified force constant tensor restores positive definiteness and translational symmetry while maintaining the crystal symmetries. Therefore, the modified tensor meets the requirements for dynamics simulations. In Sec.\ \ref{sec-III-B}, the effectiveness of this approximate method will be demonstrated through a specific example.
	
	All the first-principles calculations mentioned in this section are carried out using the Vienna \textit{ab initial} simulation package (VASP) \cite{RN293, RN294, RN295}. The force constants and phonon spectra are obtained by phonopy \cite{RN296}. The optimizations formulated in (\ref{eq-5}) are accomplished with the function \textsl{optimize.minimize} implemented in SciPy \cite{RN298}.

	\subsection{Evaluation of Damping Constants}\label{sec-II-C}
	After the construction and parameterization of Hamiltonian models, we are finally able to perform spin-lattice dynamics simulations. Before the evaluation of Gilbert damping constants, we briefly introduce the framework of SLD to cover some relevant concepts. In practice, the motion of magnetic moments follows the stochastic Landau–Lifshitz–Gilbert (SLLG) equation \cite{RN138},
	\begin{align}
		\frac{d\vec{m}_i}{dt} =& -\gamma_L \vec{m}_i \times \left(\vec{B}_i + \vec{B}_i^{fl}\right) \notag \\ 
		& -\gamma_L \alpha \frac{\vec{m}_i}{\lvert\vec{m}_i\rvert} \times \left[\vec{m}_i \times \left(\vec{B}_i + \vec{B}_i^{fl}\right)\right]
	\end{align}
	where $\gamma_L$ is the renormalized gyromagnetic ratio, $\vec{B}_i=-\partial H/\partial \vec{m}_i$ is the effective local magnetic field and $\vec{B}_i^{fl}$ refers to a stochastic field introduced by Langevin thermostat. At the same time, the motion of atoms obeys the Newton's equation,
	\begin{align}
		\frac{d\dot{u}_{i,\alpha}}{dt} = \frac{1}{M_i} \left(\vec{F}_{i,\alpha} + \vec{F}_{i,\alpha}^{fl}\right) - \nu \dot{u}_{i,\alpha}
	\end{align} 
	where $\nu$ is the damping constant and $\vec{F}_{i,\alpha}^{fl}$ refers to a stochastic force caused by thermal fluctuations. In this work, $\vec{B}_i^{fl}$ and $\vec{F}_{i,\alpha}^{fl}$ are modeled as normally distributed noises with temperature-dependent variances,
	\begin{align}
		B_{i,\beta}^{fl} &\sim N\left(0,\sqrt{2\alpha k_B T_S /\gamma \lvert\vec{m}_i\rvert \delta t}\right) \\
		F_{i,\beta}^{fl} &\sim N\left(0,\sqrt{2\nu M_i k_B T_L /\delta t}\right)
	\end{align}
	where $T_S$ and $T_L$ refer to the equilibrium temperature of spin and lattice subsystems respectively. During simulations, we can also measure the transient temperature of each subsystem with the following formulas \cite{RN252},
	\begin{align}
		T_S = \frac{\sum_i \lvert\vec{m}_i\times\vec{B}_i\rvert^2}{2k_B\sum_i\vec{m}_i\cdot \vec{B}_i},\ T_L = \frac{1}{2k_B N}\sum_{i,\alpha} M_i \dot{u}_{i,\alpha}^2
	\end{align}

	In this work, the LLG equation is numerically solved with the semi-implicit SIB method proposed by Mentink et al. \cite{RN127}. The Newton's motion equation is integrated using the Grønbech-Jensen-Farago Verlet-type method \cite{RN136}. To ensure the stability of those algorithms, a step length of $0.5$ or $0.2$ fs is adopted \cite{RN133}, where the shorter one is used in energy-conserving simulations.
	
	Based on the combination of atomistic spin dynamics (ASD) and SLD simulations, a new scheme is proposed to evaluate the damping constant in magnetic materials. Here is the basic flow of this method and more details of a specific application are presented in Sec.\ \ref{sec-III-B}. 
	\begin{enumerate}
		\item Freeze the spin degree of freedom and thermalize the lattice from $0$ to $T_L$ in the simulation.
		\item Fix atomic positions and raise the temperature of spin to $T_S > T_L$. Compared to $T_L > T_S$, this type of nonequilibrium state is more common in actual scenarios.
		\item\label{step-3} Perform an energy-conserving SLD simulation to relax the system. Normally, the spin temperature will decrease to the same as lattice and stay there till the end. 
		\item\label{step-4} Conduct a series of ASD simulations with different Gilbert damping constants. The initial states are the same as in step \ref{step-3} and the equilibrium temperatures are set to be $T_L$.
		\item\label{step-5} Compare the cooling rates $\partial T_S/\partial t$ of spin system between SLD and ASD simulations to evaluate the equivalent Gilbert damping constant contributed by spin-lattice coupling.
	\end{enumerate}

	The key point behind step \ref{step-5} is that the cooling rates observed in ASD simulations are related to the assigned damping constants, while in SLD simulation the cooling rate is determined by the strength of spin-lattice coupling. Note that the former relation can be viewed as a natural deduction of the LLG equation,
	\begin{align}
		\frac{\partial T_S}{\partial t} &= \frac{1}{C_V}\frac{\partial E_{mag}}{\partial t} \propto -\frac{1}{C_V}\left(\frac{\partial\vec{m}}{\partial t}\cdot\vec{B}\right) \notag\\ 
		&\propto -\frac{1}{C_V} \left[\left(\frac{\alpha}{m}\vec{m}\times \frac{\partial\vec{m}}{\partial t}\right)\cdot\vec{B}\right] \propto \alpha
	\end{align}
	where we have used Eq.\ (\ref{eq-1}) and simplified the formula of magnetic energy as $E_{mag}\propto-\vec{m}\cdot\vec{B}$.

	\section{results}\label{sec-III}
	This section is divided into four parts. In Sec.\ \ref{sec-III-A}, several test results are presented to validate the accuracy of SLD simulations, which are implemented in the PASP package. Subsequently, detailed calculations on three magnetic materials, namely Y$_3$Fe$_5$O$_{12}$, MnFe$_2$O$_4$ and Cr$_2$O$_3$, are discussed in the rest parts. 

	\subsection{Validations}\label{sec-III-A}
\begin{figure*}[tbp]
		\centering
		\includegraphics[width=\linewidth]{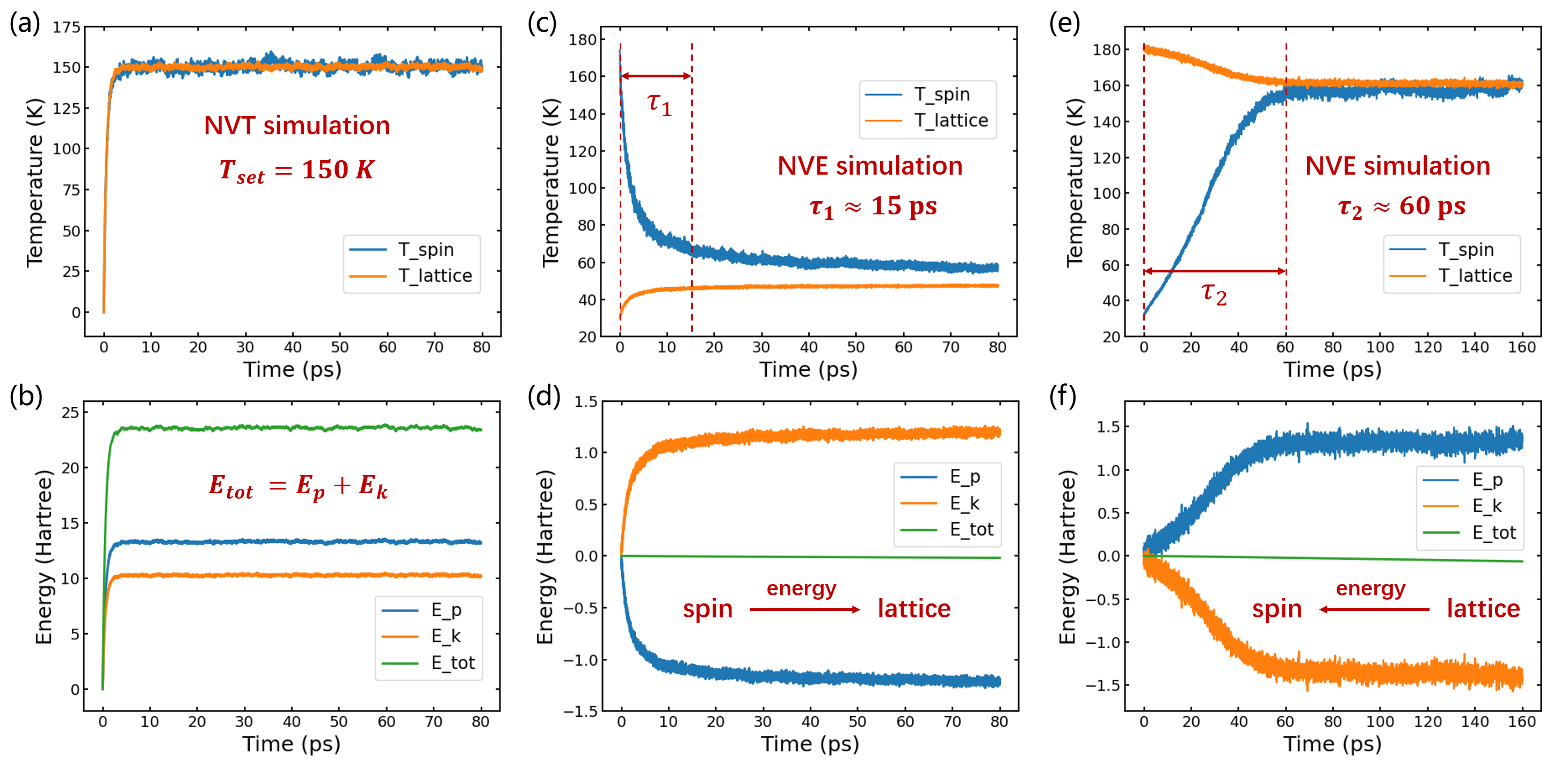}
		\caption{\label{fig-1}NVT and NVE relaxations of a spin-lattice coupled system (Cr$_2$O$_3$) within the framework of spin-lattice dynamics. The top row plots the time evolution of temperatures and the bottom row shows the variation of potential, kinetic and total energies. (a) \& (b): NVT thermalization from $T_L=T_S=0K$ to $T_L=T_S=150K$. (c) \& (d): NVE relaxation with $T_L=30K$, $T_S=175K$ initially. (e) \& (f): NVE relaxation with $T_L=180K$, $T_S=30K$ initially.}
	\end{figure*}

	In order to guarantee the reliability of our conclusions obtained from dynamics simulations, a series of pretests were carried out. We select some representative results and present them in Fig.\ \ref{fig-1}, where Cr$_2$O$_3$ is taken as the object to be studied. 
	
	Firstly, we set the ground state of Cr$_2$O$_3$ as the initial state and performed a NVT simulation with $T_{set}=150K$. As shown in Fig.\ \ref{fig-1}(a), the temperature of spin and lattice subsystems increased to $150K$ in less than 5 ps and stayed there till the end. Since we can approximate $E_k = 0.5 E_L$ and $E_p = 0.5 E_L + E_S$, Fig.\ \ref{fig-1}(b) also indicates that the contribution of phonons and magnons to the excited state energy is around 87.5\% and 12.5\% respectively. This result could be verified from another perspective. Note that there are totally 10 atoms in the unit cell of Cr$_2$O$_3$, which contribute $30k_B$ to the heat capacity. Meanwhile, the 4 magnetic atoms will contribute another $4k_B$ in the low temperature region. Therefore, we can estimate that the contribution of magnons to the total heat capacity is close to 11.8\%, which is consistent with the result from dynamics simulations.
	
	In Figs.\ \ref{fig-1}(c) \& \ref{fig-1}(d), the initial state was set to be a nonequilibrium state with $T_L=30K$ and $T_S=175K$. As we expected, the total energy was well conserved when the system evolved to equilibrium. In addition, the final temperature fell within the range of $48K\sim 55K$, which agrees with our previous analysis of the heat capacities.
	
	Lastly, we simulated the relaxation process using another nonequilibrium excited state with $T_L=180K$ and $T_S=30K$ as the initial state. As shown in Figs.\ \ref{fig-1}(e) \& \ref{fig-1}(f), the temperature of spin system increased gradually to equilibrium with the total energy conserved throughout the simulation. Also, the final temperature is around $160K$, which matches well with our analysis. It should be pointed out that there exist two notable differences between this case and the previous. Firstly, the subsystems ultimately evolved to a same temperature in a finite time, which alleviated our concerns about the accuracy of SLD simulations. Besides, the relaxation time ($\tau_2$) was much longer than that ($\tau_1$) in Fig.\ \ref{fig-1}(c). For this phenomenon, a qualitative explanation is presented below.

	Based on the theory of second quantization, the Hamiltonian model presented in Sec.\ \ref{sec-II-A} can be expressed in the following form \cite{RN154, RN145},
	\begin{align}
		H_L &= \sum_{qp} \hbar\omega_{qp} (b_{qp}^\dagger b_{qp} + 1/2) \\
		H_S &= \sum_\lambda \epsilon_\lambda a_\lambda^\dagger a_\lambda + Const.\\
		H_{SLC} &= \sum_{\lambda,qp} M_{\lambda,qp} a_{\lambda-q}^\dagger a_\lambda \left(b_{qp}^\dagger - b_{-qp}\right)
	\end{align}
	where $b_{qp}$ denotes the annihilation operator of phonons with wave vector $q$ in branch $p$, and $a_\lambda$ represents the annihilation operator of magnons with wave vector $\lambda$. All the parameters, namely $\omega_{qp}$, $\epsilon_\lambda$ and $M_{\lambda,qp}$, can be determined from the effective Hamiltonian model in principle. According to the Fermi's golden rule, we have
	\begin{widetext}
	\begin{align}
		W\left\{n_{\lambda-q},n_\lambda,N_{qp} \to n_{\lambda-q}+1,n_\lambda-1,N_{qp}+1\right\} &= \frac{2\pi}{\hbar} \lvert M_{\lambda,qp}\rvert^2 (n_{\lambda-q}+1)(n_\lambda)(N_{qp}+1) \delta(\epsilon_{\lambda-q}-\epsilon_\lambda+\hbar\omega_{qp}) \label{eq-6}\\[3mm]
		W\left\{n_{\lambda-q},n_\lambda,N_{-qp} \to n_{\lambda-q}+1,n_\lambda-1,N_{-qp}-1\right\} &= \frac{2\pi}{\hbar} \lvert M_{\lambda,qp}\rvert^2 (n_{\lambda-q}+1)(n_\lambda) (N_{-qp}) \delta(\epsilon_{\lambda-q}-\epsilon_\lambda-\hbar\omega_{-qp}) \label{eq-7}
	\end{align}
	\end{widetext}
	where $W$ represents the probability of one-phonon emission or absorption, $n_\lambda$ denotes the occupation number of magnons and $N_{qp}$ stands for phonons. Both $n_\lambda$ and $N_{qp}$ can be evaluated approximately using the Bose–Einstein distribution. According to the above formulas, the scattering rate $W$ grows linearly with $N$ and quadratically with $n$. Compared to Fig.\ \ref{fig-1}(c), there are more phonons but fewer magnons in the case of Fig.\ \ref{fig-1}(e), thus leading to a lower transition probability and a longer relaxation time. More technical details about the second quantization of interactions between phonons and magnons can be found in Ref. \cite{RN154, RN145}.
	
	\subsection{Damping constants in Y$_3$Fe$_5$O$_{12}$}\label{sec-III-B}
\begin{figure*}[tbp]
		\centering
		\includegraphics[width=\linewidth]{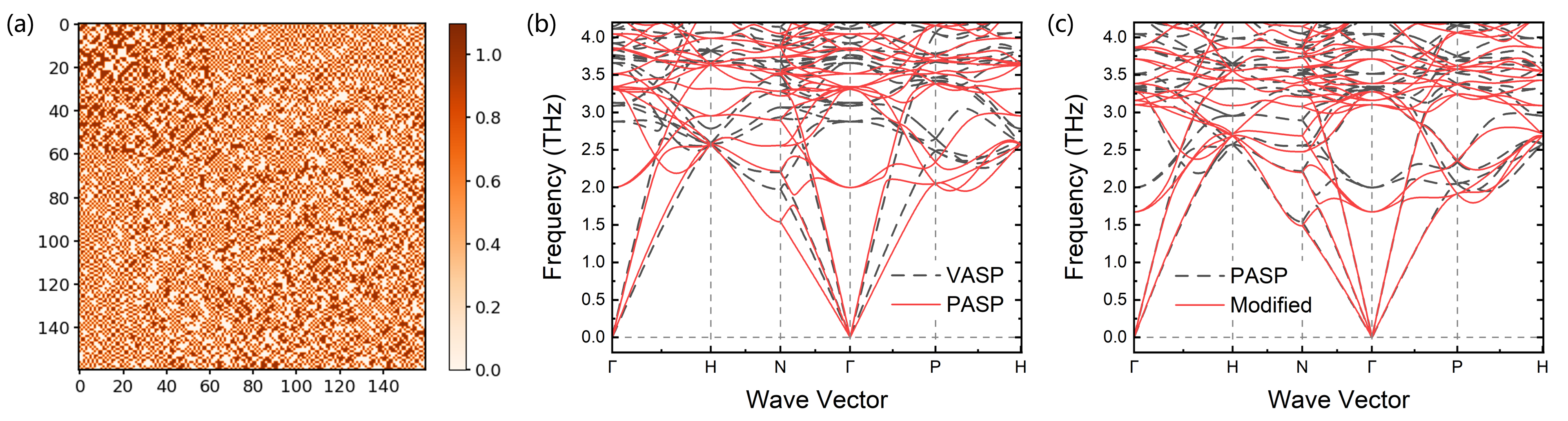}
		\caption{\label{fig-3}(a) The selection of force constant tensor elements for the cubic cell of YIG. An $160\times160$ zero-one matrix is used to show the result of selection, in which '1' denotes the interactions within cutoff radius and '0' represents the elements that are artificially eliminated. (b) The phonon spectrum calculated from the force constant tensor before and after the elimination of tensor elements. (c) The phonon spectrum calculated from the force constant tensor before and after the optimization of independent variables.}
	\end{figure*}

	In the field of spintronics, Y$_3$Fe$_5$O$_{12}$ (yttrium iron garnet, YIG) has gained much attention due to its ultra-low magnetic damping \cite{RN164, RN163, RN285}. The unique property of this material motivated us to investigate the intrinsic mechanism behind it. The crystal structure of YIG is presented in Fig.\ \ref{fig-3}(a). There are totally 80 atoms in the primitive cell, of which 12 Fe ions are located in the center of oxygen tetrahedrons while the other 8 Fe ions are sited in oxygen octahedrons. The magnetic ground state of YIG is illustrated in Fig.\ \ref{fig-3}(b). The Fe ions situated in different chemical environments contribute spins in opposite directions, which makes YIG a typical ferrimagnetic material.
	
	\begin{figure}[tbp]
		\centering
		\includegraphics[width=\linewidth]{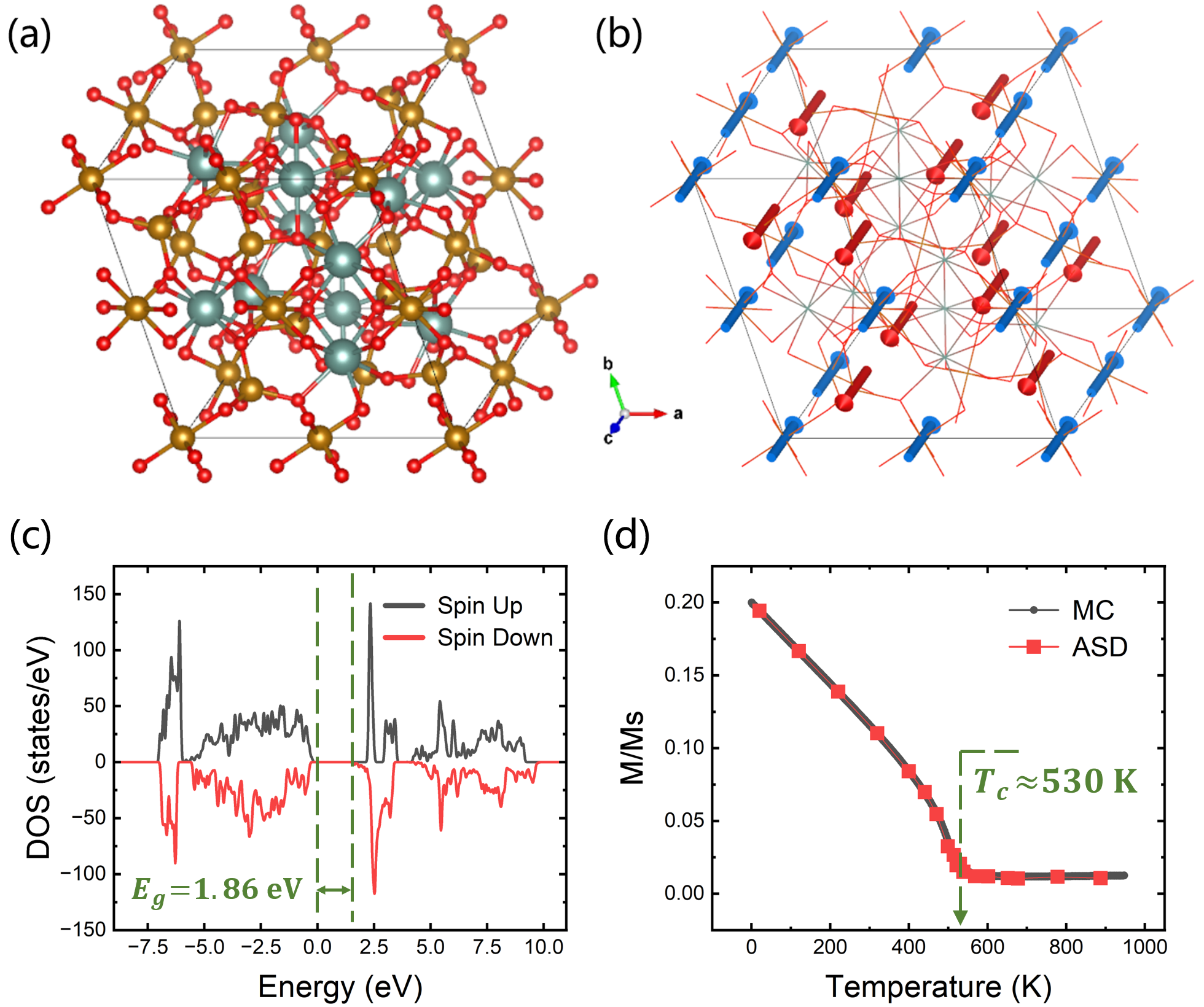}
		\caption{\label{fig-2}(a) The primitive cell of Y$_3$Fe$_5$O$_{12}$. The golden balls represent iron atoms, the cyan ball stand for yttrium atoms, and the red balls refer to oxygen atoms. (b) The magnetic ground state of YIG. The arrows of different colors represent the spin directions of Fe atoms. (c) The density of states obtained by DFT calculations. (d) The temperature dependence of average magnetization measured in MC and ASD simulations. For YIG, the phase transition point from ferrimagnetic to paramagnetic lies in 530 K approximately.}
	\end{figure}
	
	In order to evaluate the Gilbert damping constants in YIG, our first step is to prepare an effective Hamiltonian model. Considering the balance between precision and efficiency, the cutoff radius of interactions was set to be 11.0 Bohr for atomic pairs and 6.7 Bohr for 3-body clusters. After symmetry analyses, we identified 612 nonequivalent interactions in total, which included 6 Heisenberg exchange terms and 9 spin-lattice coupling terms. 
	
	To determine the interaction parameters, we carried out a series of first-principles calculations, where a cubic cell was adopted to reduce the interference between adjacent cells caused by periodic boundary conditions. Following the settings in Ref.\ \cite{RN168}, we utilized the projector augmented-wave (PAW) method \cite{RN301} and revised Perdew-Burke-Ernzerhof exchange-correlation functional for solids (PBEsol) \cite{RN302} in our calculations. Besides, the DFT+U method in its simplified form \cite{RN303} was employed where the effective Hubbard U parameter was set to be 4 eV for the 3$d$ electrons of Fe ions. In addition, a cutoff energy of 520 eV for plane wave basis and a $\Gamma$-centered $2\times2\times2$ mesh of $k$-points were used in the DFT calculations. 
	
	In Figure \ref{fig-2}(c), we present the density of states (DOS) for YIG. With a band gap of 1.863 eV, there is hardly any electric current occurring in the low temperature region. Moreover, the Heisenberg exchange coefficients of YIG is listed in Table \ref{tab-1}. To verify the accuracy of these parameters, we conducted both Monte Carlo (MC) and ASD simulations. The temperature dependence of average magnetization is shown in Fig.\ \ref{fig-2}(d), which reveals the critical temperature of YIG to be 530 K. This result is slightly lower than the measured Curie temperature, $T_C=560K$ \cite{RN164}, but falls within our tolerance. The calculated results of coupling constants are provided in the supplementary material.
	
	\begin{table}[tbp]
		\renewcommand{\arraystretch}{1.5}
		\caption{\label{tab-1}The Heisenberg exchange coefficients J of YIG, where an effective spin $S=1$ is adopted. For the Fe$^O-$Fe$^O$ pairs, the Greek letters ($\alpha$ \& $\beta$) refer to different chemical environments. All the results are calculated with the four-state method.}
		\begin{tabular}{p{10em}<{\centering} p{9em}<{\centering} p{6.3em}<{\centering}}
			\hline
			\hline
			Spin Pair. & Distance (Angst) & J (meV) \\ 
			\hline
			1NN Fe$^T-$Fe$^O$ & 3.445 & 47.414 \\ 
			1NN Fe$^T-$Fe$^T$ & 3.774 & 2.399 \\ 
			1NN Fe$^O-$Fe$^O$ ($\alpha$) & 5.337 & 0.538 \\ 
			1NN Fe$^O-$Fe$^O$ ($\beta$) & 5.337 & 5.055 \\ 
			2NN Fe$^T-$Fe$^O$ & 5.555 & 0.285 \\ 
			2NN Fe$^T-$Fe$^T$ & 5.765 & 3.437 \\ 
			\hline 
			\hline
		\end{tabular}
	\end{table}
	
	Next, we come to deal with the force constant tensor. In order to demonstrate the impact of locality and validate the effectiveness of our optimization method, we present some results pertaining to the tensor of YIG in Table \ref{tab-2}. Here we use ``VASP'' to tag the original tensor obtained from DFT calculations, ``PASP'' to label the modified tensor in which interactions beyond the cutoff radius are eliminated, and ``Modified'' to label the tensor after optimization of independent variables. As shown in Table \ref{tab-2}, the ``PASP'' tensor violated the acoustic sum rule and was not positive semi-definite, whereas these issues were resolved for the ``Modified'' tensor. Although an obvious difference existed between the ``PASP'' and ``Modified'' tensor in terms of their eigenvalues, we still assumed the target system could be reasonably described by the ``Modified'' tensor and the validity of this assumption would be verified by the calculated results of damping constants. Additional details regarding the selection of tensor elements and the deviation of phonon spectra are provided in Fig.\ \ref{fig-3}. According to figure \ref{fig-3}(b) and \ref{fig-3}(c), the major deviation in phonon spectra resulted from the elimination of tensor elements, rather than the subsequent modification. 
	
	\begin{table}[tbp]
		\renewcommand{\arraystretch}{1.5}
		\caption{\label{tab-2}The force constant tensor of YIG. The columns labeled by A represent the sorted absolute values of $\sum_i K_{ij,\alpha\beta}$ and the columns labeled by B list the sorted eigenvalues of $K_{ij,\alpha\beta}$. For the cubic cell of YIG, we obtained the original tensor with the VASP package. Then, we eliminated the elements that represent interactions beyond the cutoff radius. This step was done by PASP. Finally, the tensor was modified to meet the requirement of translational symmetry through the optimization formulated in (\ref{eq-5}).}
		\begin{tabular}{p{2em} p{3.6em}<{\centering} p{3.6em}<{\centering} p{3.6em}<{\centering} p{3.6em}<{\centering} p{3.6em}<{\centering} p{3.6em}<{\centering}}
			\hline
			\hline
			& \multicolumn{2}{c}{\text{VASP}} & \multicolumn{2}{c}{\text{PASP}} & \multicolumn{2}{c}{\text{Modified}} \\
			\cmidrule(r){2-3} \cmidrule(r){4-5} \cmidrule(r){6-7}
			\text{No. } & \text{A} & \text{B} & \text{A} & \text{B} & \text{A} &  \text{B} \\ 
			\hline
			1 & 0.000 & 0.000  & 1.587 & -0.102 & 0.000 & 0.000 \\ 
			2 & 0.000 & 0.000  & 1.587 & -0.102 & 0.000 & 0.000 \\ 
			3 & 0.000 & 0.000  & 1.587 & -0.102	& 0.000 & 0.000 \\ 
			4 & 0.000 & 1.065  & 1.587 &  0.643 & 0.000 & 0.444 \\ 
			5 & 0.000 & 1.065  & 1.587 &  0.643 & 0.000 & 0.444 \\ 
			6 & 0.000 & 1.065  & 1.587 &  0.643 & 0.000 & 0.444 \\ 
			\hline 
			\hline
		\end{tabular}
	\end{table}

\begin{figure*}[tbp]
		\centering
		\includegraphics[width=\linewidth]{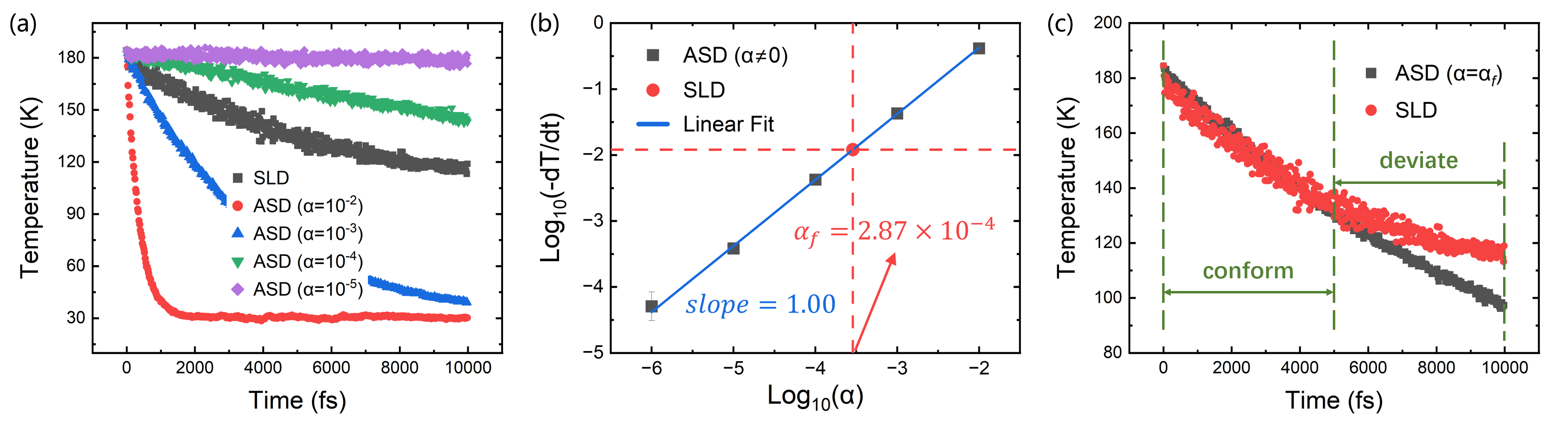}
		\caption{\label{fig-4}(a) The time evolution of spin temperature in SLD and ASD simulations. The gray line represents the SLD simulation while the others refer to the ASD simulations with different damping constants. (b) The initial cooling rates $\partial T_S/\partial t\rvert_{t=0}$ with respect to the damping constants $\alpha$, where the scaling of axis is set to be logarithm. The gray squares refer to the results of ASD simulations and the blue line acts as the linear regression. The red circle is plotted by intersection of the blue line and the horizontal red dash line, which represents the initial cooling rate in the SLD simulation. Then we can obtain the equivalent damping constant from the abscissa of the red circle. (c) The comparison between ASD and SLD simulations. In the ASD simulation, the Gilbert damping constant is set to be $\alpha=2.87\times 10^{-4}$, which is exactly the result of our evaluation from the SLD simulation.}
	\end{figure*}

	Completing the preparation of Hamiltonian model, we applied the scheme proposed in Sec.\ \ref{sec-II-C} to our first object, Y$_3$Fe$_5$O$_{12}$. An instance is presented in Figure \ref{fig-4}. We set $T_L=30K$, $T_S=180K$ for the initial nonequilibrium state and adopted an expanded supercell which contains 12800 atoms in the simulation. Fig.\ \ref{fig-4}(a) shows the time evolution of spin temperature in different types of simulations. By comparing the curves, we could roughly estimate that the equivalent damping constant in SLD simulation fell within the range of $10^{-3}\sim10^{-4}$. To make the estimation more precise, we calculated the initial cooling rates $\partial T_S/\partial t\rvert_{t=0}$ through polynomial (or exponential) fittings and plotted them in Fig.\ \ref{fig-4}(b). Afterwards, a linear regression was performed to determine the quantitative relation between $\lg(-\partial T_S/\partial t\rvert_{t=0})$ and $\lg(\alpha)$. As we expected, the cooling rates in ASD simulations were proportional to the assigned damping constants. Then, we combined the results of SLD and ASD simulations to evaluate the equivalent damping constant. This step was accomplished by identifying the intersection of red and blue lines in Figure\ \ref{fig-4}(b). Finally, the damping constant was determined to be $\alpha_f=\left(2.87\pm 0.13\right)\times 10^{-4}$ in this case. To verify our method and result, we present a comparison between SLD and ASD (where we set $\alpha=\alpha_f$) simulations in Fig.\ \ref{fig-4}(c). The curves agree well with each other in the initial stage but deviate in the second half. This phenomenon is within our expectation, because in the SLD simulation the lattice heats up as the spin cools down, thereby slowing the energy transfer between two subsystems.
	
	In addition to the above case, we have measured the equivalent damping constants under different conditions to investigate the temperature dependence of magnetic damping. The final results are summarized in Figure \ref{fig-5}. Details about the estimation of uncertainties are given in the supplementary material. For Y$_3$Fe$_5$O$_{12}$, the damping constants at different temperatures stay on the order of $10^{-4}$, which is in good agreement with the experimental results ($3.2\times10^{-4}$ \cite{RN202}, $2.2\times10^{-4}$ \cite{RN205}, $1.2$--$1.7\times10^{-4}$ \cite{RN204}). For example, the damping constant in bulk YIG was reported as $0.4\times10^{-4}$ in Ref.\ \cite{roschmann1983}. Meanwhile, our calculations yielded $\alpha = (2.8\pm0.3) \times10^{-5}$ at $\Delta T=15$ K and $\alpha = (7.0\pm0.7) \times10^{-5}$ at $\Delta T=30$ K, where both $T_L=0$ K. Therefore, the experimental value corresponds roughly to the temperature region of $\Delta T=15\sim30$ K in our study. We believe such extent of thermal excitation is quite common in all kinds of spintronics experiments. Moreover, Fig.\ \ref{fig-5} indicates that $\alpha$ is approximately proportional to the temperature difference between subsystems. This outcome is also consistent with some computational works in the past \cite{RN267, RN269}. By comparing the subfigures in Figure \ref{fig-5}, we found that $\alpha$ has little dependence on the lattice temperature, although here $T_L$ could be viewed to some extent as the ambient temperature of the spin system.
	
	\begin{figure*}[tbp]
		\centering
		\includegraphics[width=\linewidth]{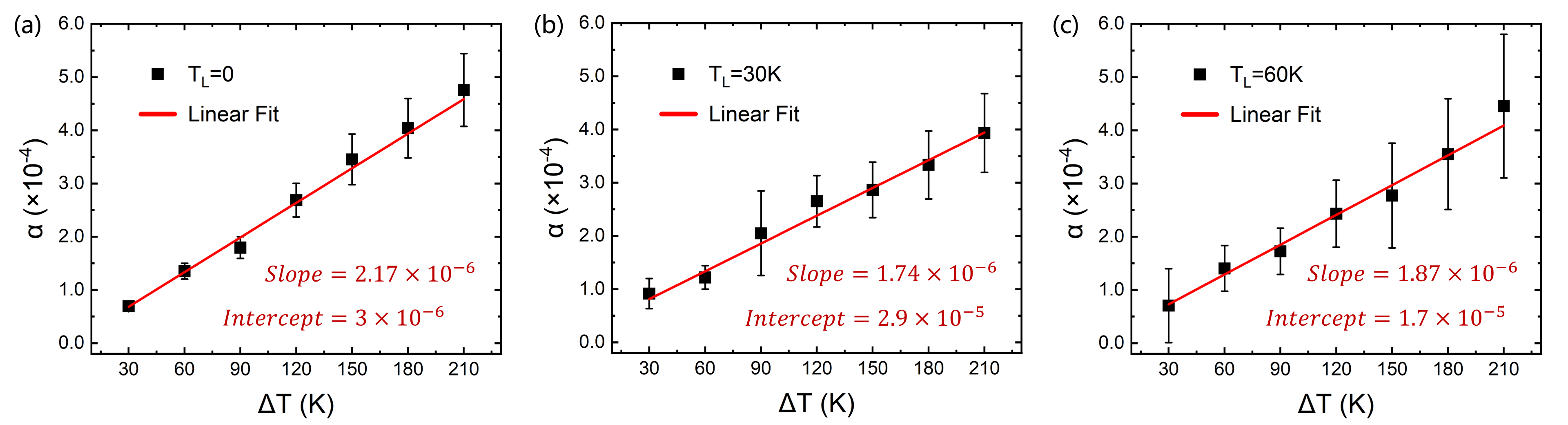}
		\caption{\label{fig-5}The temperature dependence of Gilbert damping constants for  Y$_3$Fe$_5$O$_{12}$. The label of abscissa axis $\Delta T$ refers to $T_S-T_L$ of the initial state in dynamical simulations. Measurements on the magnetic damping are performed under different initial conditions of the lattice temperature: (a) $T_L=0$, (b) $T_L=30K$, (c) $T_L=60K$.}
	\end{figure*}
	
	As a supplement to Sec.\ \ref{sec-III-A}, we further validate our simulations by analyzing the measured cooling rates in Fig.\ \ref{fig-5}(a). By subtracting Eq.\ (\ref{eq-7}) from Eq.\ (\ref{eq-6}), the transfer rate of energy between magnon and phonon systems can be expressed as,
	\begin{align}
		\dot{Q} = \sum_{qp}\hbar\omega_{qp}\langle\dot{N}_{qp}\rangle =\sum_{\lambda,qp}T_{\lambda,qp}
	\end{align}
	where $T_{\lambda,qp}$ denotes different transfer channels,
	\begin{align}
		T_{\lambda,qp} \propto  (n_\lambda-n_{\lambda-q}) N_{qp} + n_{\lambda-q} n_\lambda +1
	\end{align}
	According to the Bose–Einstein distribution, the number of magnons and phonons can be expressed as, 
	\begin{align}
		n_\lambda = \frac{1}{e^{\epsilon_\lambda/k_BT_S}-1},\ 
		N_{qp} = \frac{1}{e^{\hbar\omega_{qp}/k_BT_L}-1}
	\end{align}
	When $T_S$ is high enough and $T_L$ is close to zero, we can approximate $ n_\lambda = k_BT_S/\epsilon_\lambda\propto T_S$ and $N_{qp}$ close to zero. Under these conditions, we have $\dot{Q}\propto T_S^2$. This relation is well verified by linear regressions and the details are provided in the supplementary material.
	
	Furthermore, the accuracy of our simulations can also be proved from another perspective. According to Eqs.\ (\ref{eq-6}) and (\ref{eq-7}), the scattering rate $W$ grows quadratically with the coupling parameters $M_{\lambda, qp}$. Based on the theory of second quantization, $M_{\lambda, qp}$ shall be proportional to the coupling constants $\partial J_{ij}/\partial u_{k,\alpha}$. Therefore, under a definite condition of temperature, we have:
	\begin{align}
		\alpha \propto \dot{Q} \propto \Delta W \propto M_{\lambda, qp}^2 \propto (\partial J_{ij} / \partial u_{k,\alpha})^2
	\end{align}

\begin{figure}[tbp]
		\centering
		\includegraphics[width=0.7\linewidth]{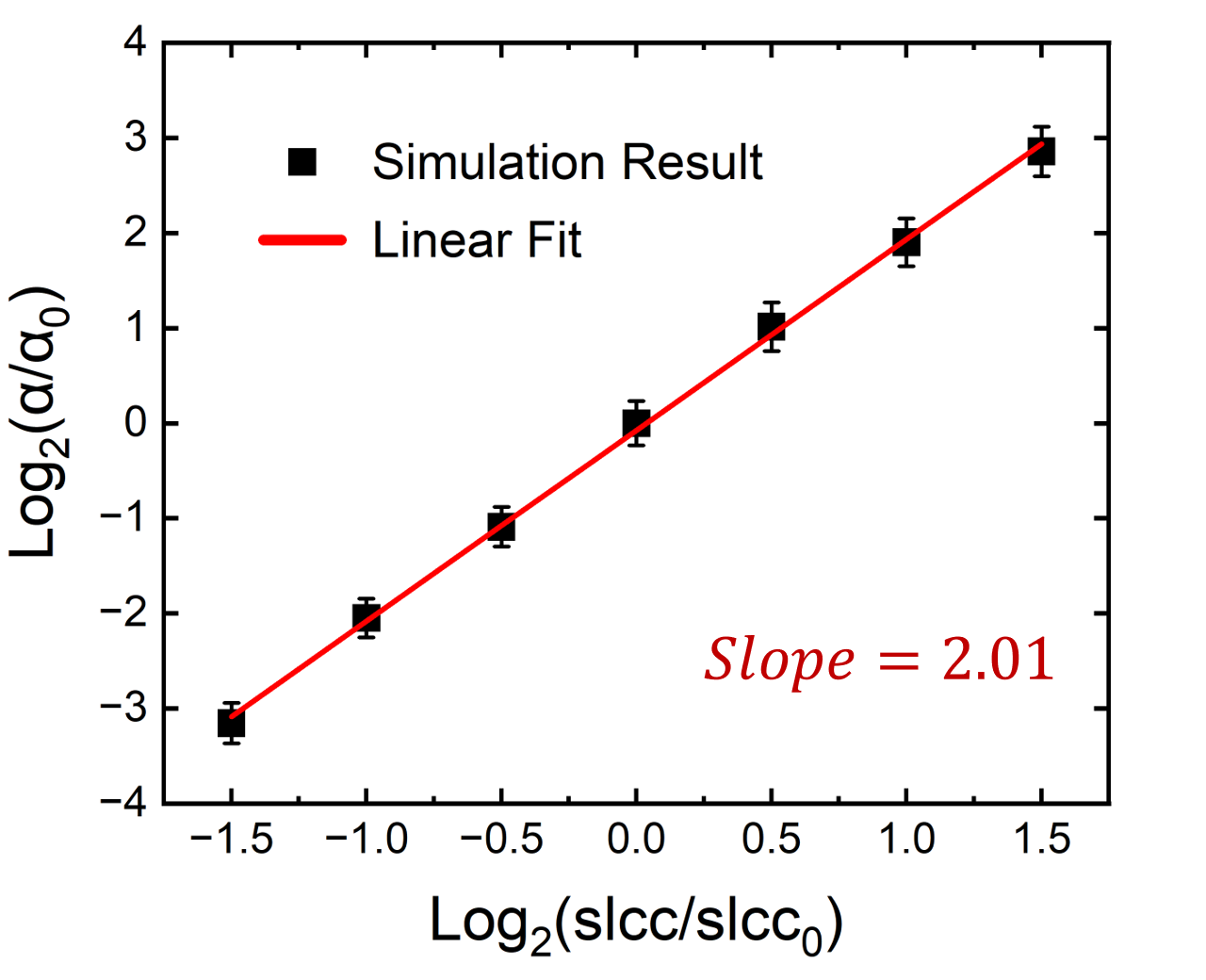}
		\caption{\label{fig-10}The relation between damping constants $\alpha$ and spin-lattice coupling constants $\partial J_{ij}/\partial u_{k,\alpha}$ in YIG. Through a linear fitting, the slope is determined to be $2.01$, which agrees well with our theoretical predictions.}
	\end{figure}

	In order to verify this relation, we adjusted the spin-lattice coupling constants of YIG coherently while keeping the other model parameters unchanged. Then, SLD simulations were carried out to evaluate the corresponding damping constants. The result is plotted in Fig.\ \ref{fig-10}, where the x-label ``slcc'' stands for the spin-lattice coupling constants and the subscript ``0'' refers to the original situation. From a linear fitting, the slope is determined to be $2.01$, which agrees well with our prediction.
	
	\subsection{Damping constants in MnFe$_2$O$_4$}\label{sec-III-C}
\begin{figure}[tbp]
		\centering
		\includegraphics[width=\linewidth]{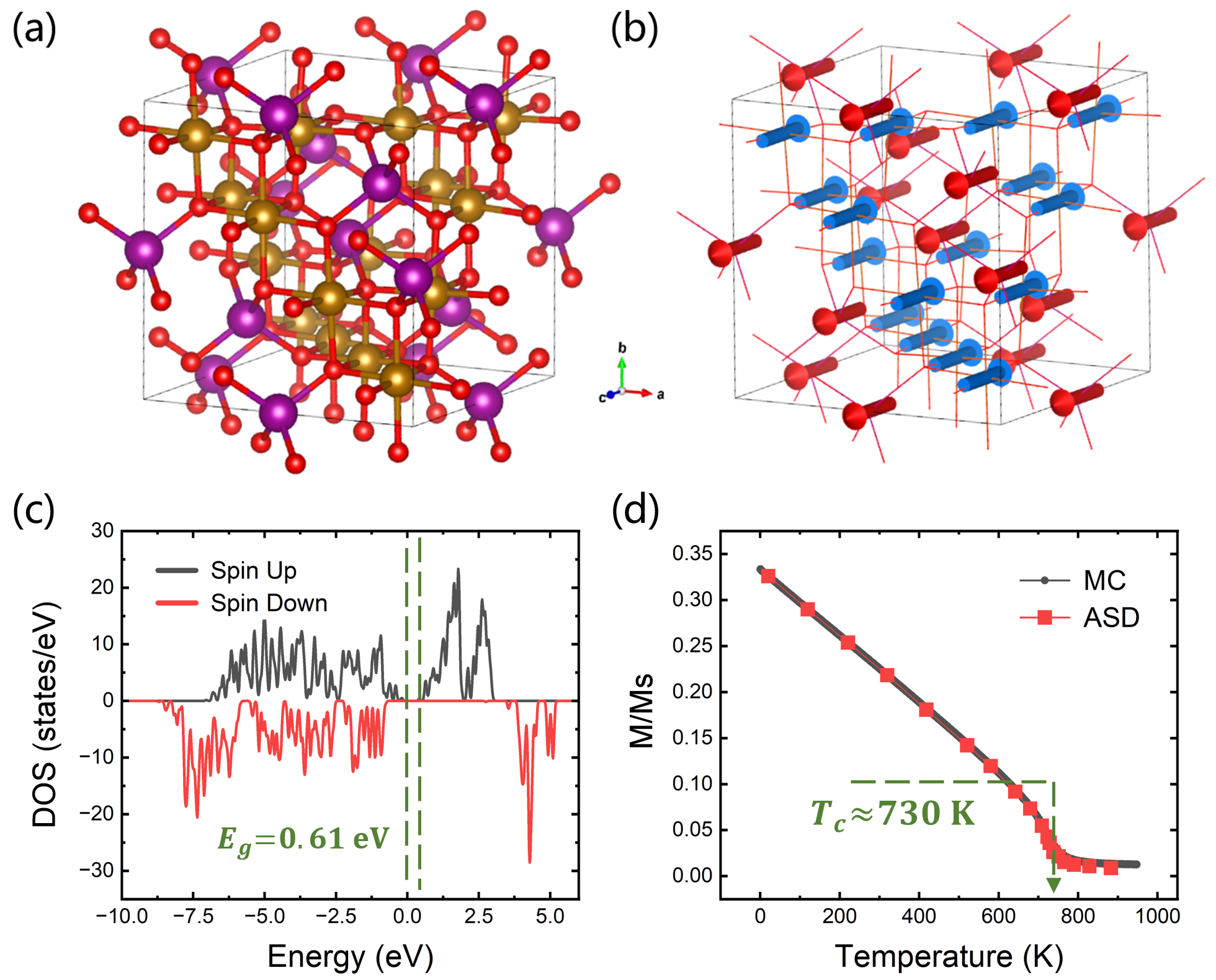}
		\caption{\label{fig-6}(a) The cubic cell of MnFe$_2$O$_4$. The purple balls represent manganese atoms, the golden balls refer to iron atoms, and the red balls stand for oxygen atoms. (b) The magnetic ground state of MFO. The arrows of different colors represent the spin directions of Mn and Fe atoms separately. (c) The density of states obtained by DFT calculations. (d) The temperature dependence of average magnetization measured in MC and ASD simulations. For MnFe$_2$O$_4$, the phase transition point from ferrimagnetic to paramagnetic lies in 730K approximately. }
	\end{figure}

	After the calculation on YIG, we applied our method to MnFe$_2$O$_4$ (MFO), which was reported to possess a large Gilbert damping constant in the literature \cite{goodenough1970, RN209}. As shown in Fig.\ \ref{fig-6}(a), MnFe$_2$O$_4$ has a typical structure of spinels, where A sites are surrounded by four oxygen atoms and B sites are located in octahedrons. Generally, spinels can be classified into normal and inverse structures according to the distribution of divalent and trivalent cations between A/B sites. In experiments, MFO usually crystallizes into a mixed phase where the normal structure occupies the major part (80\% in bulk MFO \cite{RN232}). Here, we only considered its normal structure in this work. Also, the magnetic ground state of MFO is shown in Fig.\ \ref{eq-6}(b), where the magnetic moments are antiparallel between A/B sites.
	
	Firstly, we started to construct an effective Hamiltonian model for MFO. With the same cutoff settings for YIG, we found 105 nonequivalent interactions, including 4 Heisenberg exchange terms and 10 spin-lattice coupling terms. Subsequently, DFT calculations were carried out to determine the interaction parameters. In these calculations, we adopted a cubic cell containing 56 atoms and a $\Gamma$-centered $4\times4\times4$ grid mesh in the reciprocal space. Besides, $U_{\text{Mn}}=3.3$ eV and $U_{\text{Fe}}=3.6$ eV were used as the effective Hubbard parameters \cite{RN232}. With the exception of aforementioned settings, all the relevant first-principles calculations were performed under the same conditions as in Sec.\ \ref{sec-III-B}.
	
	The DOS of MnFe$_2$O$_4$ is plotted in Fig.\ \ref{fig-6}(c), yielding a calculated band gap of 0.612 eV. This value does not match with the result of transport experiments, which reported a much smaller band gap ($0.04$--$0.06$ eV) \cite{RN238}. In addition, MC and ASD simulations were performed using the Heisenberg exchange coefficients listed in Table \ref{tab-3}. The temperature dependence of average magnetization, shown in Fig.\ \ref{fig-6}(d), suggests the critical temperature to be around 730 K. This result is significantly higher than the measured value of 573 K \cite{RN235}. Both of the above discrepancies may be attributed to the inevitable difference between the ideal normal spinel structure in calculations and the partially disordered samples in reality. Despite this problem, we proceeded to describe the target system with our Hamiltonian model and expected to see how far the calculated results of damping constants would differ from experimental values. 
	
	\begin{table}[tbp]
		\renewcommand{\arraystretch}{1.5}
		\caption{\label{tab-3}The exchange coefficients J of MnFe$_2$O$_4$, where an effective spin $S=1$ is adopted.}
		\begin{tabular}{p{9em}<{\centering} p{9em}<{\centering} p{7.3em}<{\centering}}
			\hline
			\hline
			Spin Pair. & Distance (Angst) & J (meV) \\ 
			\hline
			1NN Fe-Fe & 3.003 & 6.835 \\ 
			1NN Mn-Fe & 3.521 & 33.224 \\ 
			1NN Mn-Mn & 3.667 & 3.956 \\ 
			2NN Fe-Fe & 5.201 & 0.929 \\ 
			\hline 
			\hline
		\end{tabular}
	\end{table}

\begin{figure*}[tbp]
		\centering
		\includegraphics[width=\linewidth]{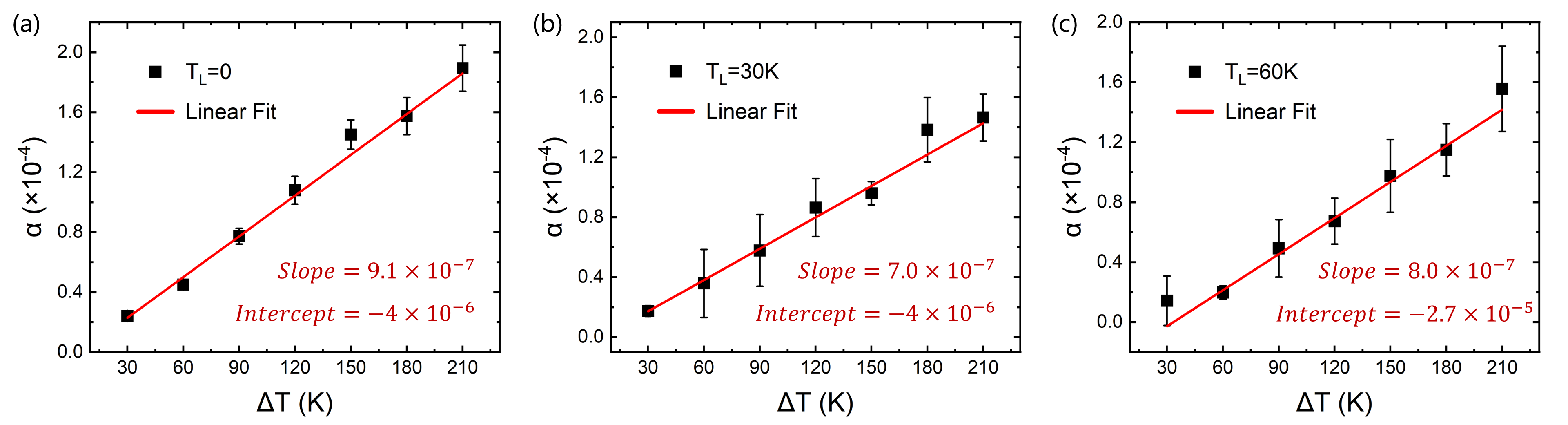}
		\caption{\label{fig-7}The temperature dependence of Gilbert damping constants for MnFe$_2$O$_4$. The label of abscissa axis $\Delta T$ refers to $T_S-T_L$ of the initial state in dynamical simulations. Measurements on the magnetic damping are performed under different initial conditions of the lattice temperature: (a) $T_L=0$, (b) $T_L=30K$, (c) $T_L=60K$.}
	\end{figure*}

	After the preparation of Hamiltonian model, we conducted dynamics simulations to evaluate the equivalent damping parameters in MFO at different temperatures. A supercell containing 13440 atoms was adopted in the simulation, and the results are summarized in Fig.\ \ref{fig-7}. The average of calculated damping constants is around $8\times 10^{-5}$, which is much smaller than the measured value, $1.0 \times 10^{-2}$ \cite{goodenough1970, RN209}. Two factors may account for this inconsistency. Firstly, the inhomogeneity in real MnFe$_2$O$_4$ samples greatly enhances the scattering of magnons and phonons, thereby increasing the damping constants. Additionally, due to the narrow band gap observed in experiments, eddy currents can arise at finite temperatures, which leads to a rapid loss of energy in the form of joule heat. As the result of these factors, we failed to obtain a reasonable estimation of Gilbert damping constants for MnFe$_2$O$_4$ with our methodology. On the other side, the contribution of different relaxation mechanisms to FMR linewidth has been studied comprehensively for MnFe$_2$O$_4$ in Ref.\ \cite{RN238}, which further confirms our analyses. 

	\subsection{Damping constants in Cr$_2$O$_3$}\label{sec-III-D}
\begin{figure}[tbp]
		\centering
		\includegraphics[width=\linewidth]{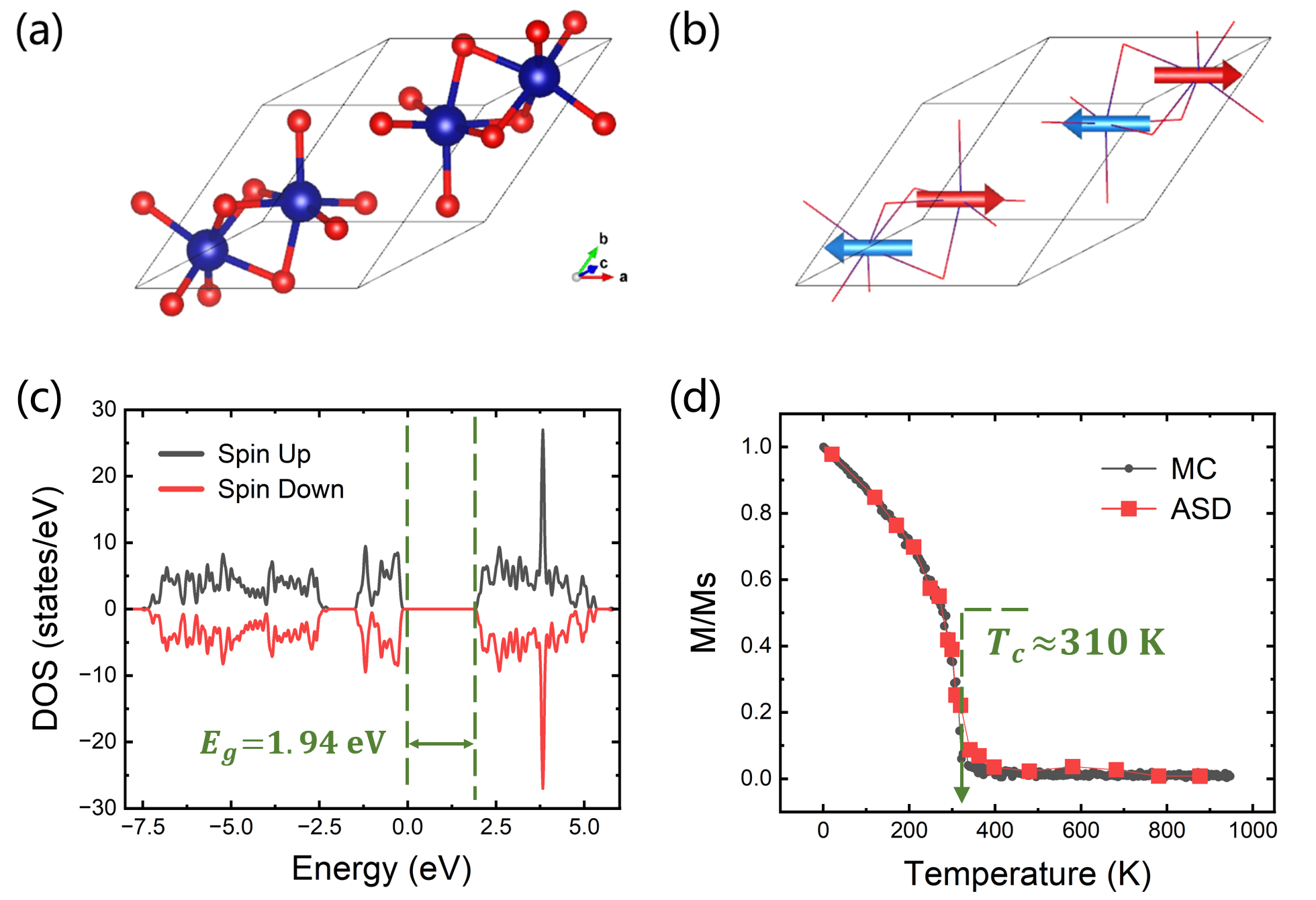}
		\caption{\label{fig-8}(a) The primitive cell of Cr$_2$O$_3$. The dark blue balls represent chromium atoms, and the red balls stand for oxygen atoms. (b) The magnetic ground state. The arrows of different colors represent the spin directions of Cr atoms. (c) The density of states obtained by DFT calculations. (d) The temperature dependence of sublattice magnetization measured in MC and ASD simulations. For Cr$_2$O$_3$, the phase transition point from ferrimagnetic to paramagnetic lies in 310K approximately. }
	\end{figure}

	Chromia (Cr$_2$O$_3$) is a well-known collinear magnetoelectric antiferromagnet, which holds great prospects in the field of spintronics \cite{RN225, RN257, makushko2022}. As shown in Fig.\ \ref{fig-8}(a), the primitive cell of Cr$_2$O$_3$ contains 10 atoms, with each chromium atom bonded to the six oxygen atoms around it. Additionally, Fig.\ \ref{fig-8}(b) displays the magnetic ground state of Cr$_2$O$_3$, where the spins of two nearest neighboring Cr atoms are oriented in opposite directions.
	
	As a preliminary step in constructing the Hamiltonian model, we set the cutoff radius of interactions to be 11.0 Bohr for atomic pairs and 7.0 Bohr for 3-body clusters. Through symmetry analyses, we identified 319 nonequivalent interactions, including 5 Heisenberg exchange terms and 21 spin-lattice coupling terms. 
	
	Afterwards, a series of first-principles calculations were performed to determine the model parameters. Following the settings in Ref.\ \cite{RN226}, we adopted a hexagonal cell of Cr$_2$O$_3$ which contained a total of 90 atoms in the calculations. Additionally, we used the LSDA+U method in its full spherically symmetric form \cite{RN304}. As to the Hubbard parameters, $J$ was fixed at its recommended value of 0.6 eV, and $U$ was adjusted to fit the Néel temperature observed in experiments \cite{brown1969magneto}. We found $U=2.0$ eV was the optimal value for 3$d$ electrons of Cr ions. Except for the settings specified above, all the DFT calculations were conducted under the same conditions as in Sec.\ \ref{sec-III-C}.
	
	The DOS of Cr$_2$O$_3$ is plotted in Fig.\ \ref{fig-8}(c), which yields a calculated band gap of 1.935 eV. This value indicates that the energy dissipation of electric currents can be neglected in this system. Additionally, we list the Heisenberg exchange coefficients of chromia in Table \ref{tab-4}. Both MC and ASD simulations were performed to investigate the temperature dependence of sublattice magnetization. According to Fig.\ \ref{fig-8}(d), the critical point was determined to be 310 K approximately, which was quite consistent with experimental observations. Also, the force constants of Cr$_2$O$_3$ went through the modification formulated in Sec.\ \ref{sec-II-B}, and the spin-lattice coupling parameters are provided in the supplementary material. 

	\begin{table}[tbp]
		\renewcommand{\arraystretch}{1.5}
		\caption{\label{tab-4}The exchange coefficients J of Cr$_2$O$_3$, in which an effective spin $S=1$ is adopted.}
		\begin{tabular}{p{9em}<{\centering} p{9em}<{\centering} p{7.3em}<{\centering}}
			\hline
			\hline
			Spin Pair. & Distance (Angst) & J (meV) \\ 
			\hline
			1NN Cr-Cr & 2.640 &  44.778 \\
			2NN Cr-Cr & 2.873 &  29.269 \\
			3NN Cr-Cr & 3.411 & -0.182 \\
			4NN Cr-Cr & 3.635 &  0.007 \\
			5NN Cr-Cr & 4.137 & -0.500 \\
			\hline 
			\hline
		\end{tabular}
	\end{table}
	
\begin{figure*}[tbp]
		\centering
		\includegraphics[width=\linewidth]{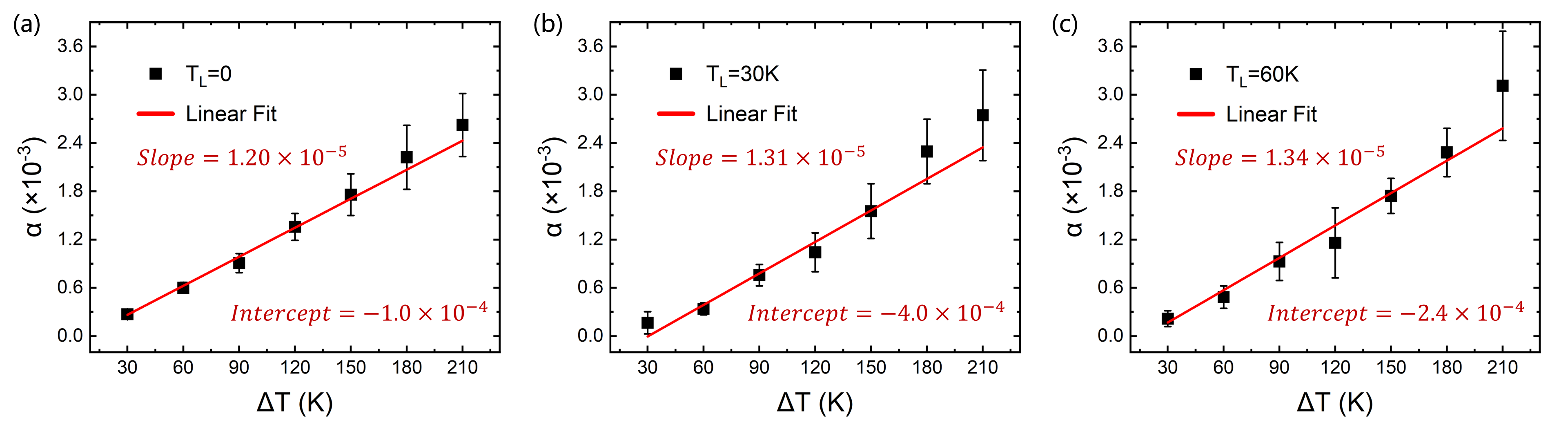}
		\caption{\label{fig-9}The temperature dependence of Gilbert damping constants for Cr$_2$O$_3$. The label of abscissa axis $\Delta T$ refers to $T_S-T_L$ of the initial state in dynamical simulations. Measurements on the magnetic damping are performed under different initial conditions of the lattice temperature: (a) $T_L=0$, (b) $T_L=30K$, (c) $T_L=60K$.}
	\end{figure*}

	After the construction of Hamiltonian model, we conducted a series of dynamics simulations to evaluate the equivalent damping parameters in Cr$_2$O$_3$. An expanded hexagonal cell containing 14400 atoms was adopted for the simulation, and the results are summarized in Fig.\ \ref{fig-9}. As two specific cases, our calculation yielded $\alpha = (1.31\pm0.14) \times 10^{-4}$ at $\Delta T=15$ K and $\alpha = (2.7\pm0.3) \times 10^{-4}$ at $\Delta T=30$ K, where both $T_L=0$ K. Therefore, the calculated damping constants within $\Delta T=15\sim30$ K are quite close to $2\times10^{-4}$, which is the estimated value reported in Ref.\ \cite{RN222}. 
	
	Furthermore, the damping constants in Cr$_2$O$_3$ exhibit a significant non-linear relation with the temperature difference of subsystems. Through logarithmic fittings, we calculated the power exponents for Figures \ref{fig-9}(a) to \ref{fig-9}(c), and the results were 1.17, 1.62, 1.38. If we disregard the difference between $\Delta T$ and $T$ for the moment, these values are in good agreement with the theoretical prediction of Kasuya and LeCraw \cite{RN306}. According to their study, the relaxation rate varies as $T^{n}$ where $n=1\sim2$ while $n=2$ corresponds to a larger regime of temperatures.
		
\begin{figure}[tbp]
		\centering
		\includegraphics[width=\linewidth]{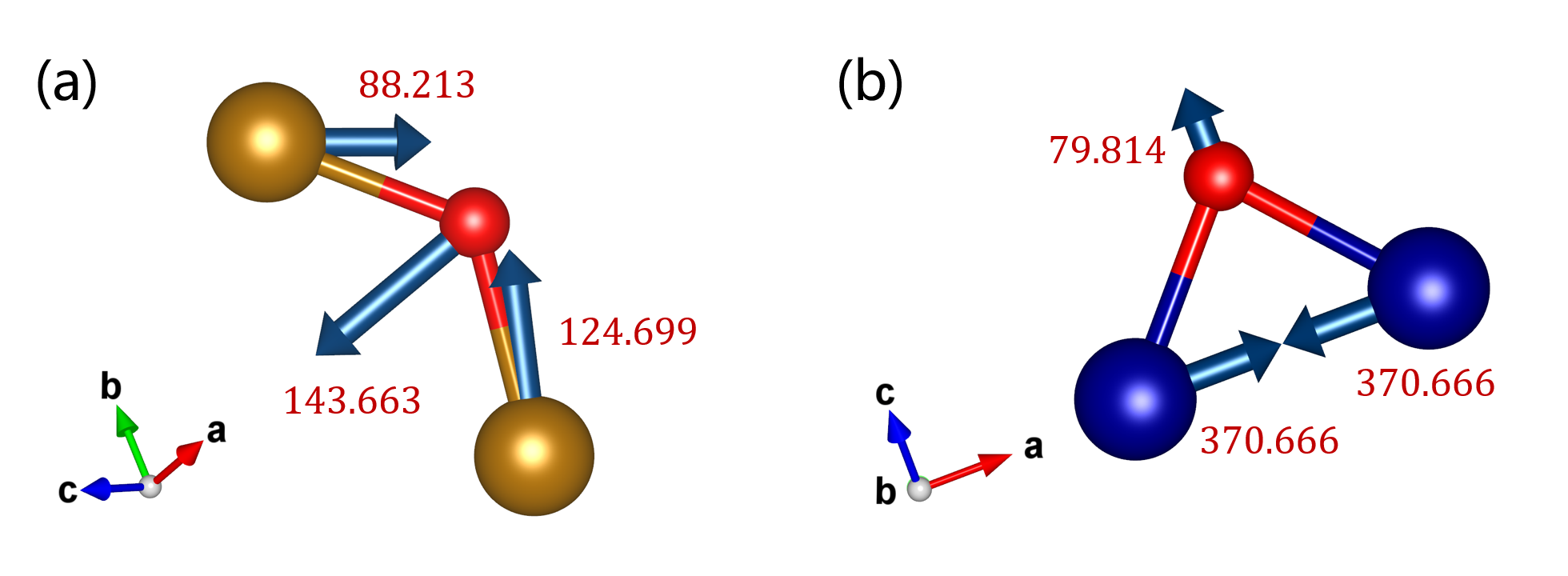}
		\caption{\label{fig-11}(a) The 1NN Fe$^T$-Fe$^O$ pair in Y$_3$Fe$_5$O$_{12}$. (b) The 1NN Cr-Cr pair in Cr$_2$O$_3$. The steel blue arrow stands for the orientation of $\partial J / \partial u$ and the red number along with it represents the magnitude in unit of meV/Angst.}
	\end{figure}

	Compared to YIG, the greater magnetic damping observed in chromia can be attributed to its significantly stronger spin-lattice coupling. As shown in Fig.\ \ref{fig-11}, the magnitude of principal spin-lattice coupling constant in Cr$_2$O$_3$ is two or three times larger than that in YIG. This could be explained by the fact that direct exchange interaction between two magnetic atoms decreases rapidly with their distance \cite{WOS005}. Therefore, owing to the shorter distance of Cr-Cr pair, the direct exchange interaction between neighboring Cr atoms is believed to have a great contribution to the spin-lattice coupling in Cr$_2$O$_3$.

	\section{conclusions}\label{sec-IV}
	In summary, we propose a scheme to evaluate the contribution of spin-lattice coupling to the Gilbert damping in insulating magnetic materials. Our methodology involves first-principles based Hamiltonian models and spin-lattice dynamics simulations. Following a series of validations, we applied our method to three magnetic materials, namely Y$_3$Fe$_5$O$_{12}$, MnFe$_2$O$_4$ and Cr$_2$O$_3$. Their damping constants were estimated separately, and the results show that, in general, $\alpha$ is approximately proportional to the temperature difference between spin and lattice subsystems. Under the condition of $\Delta T=30$ K, the calculated damping constants are averaged to be $0.8\times10^{-4}$ for YIG, $0.2\times10^{-4}$ for MFO and $2.2\times 10^{-4}$ for Cr$_2$O$_3$. The results for YIG and Cr$_2$O$_3$ are in good agreement with experimental measurements, while the discrepancy for MFO can be attributed to the inhomogeneity and small band gap in real samples. Overall, the approach presented in this work holds great promise for accurately predicting the Gilbert damping constants for magnetic insulators.
	
	\begin{acknowledgments}
		This work is supported by the National Key R\&D Program of China (No. 2022YFA1402901 ), the National Natural Science Foundation of China (Grant Nos. 11825403, 11991061, and 12188101), the Guangdong Major Project of the Basic and Applied Basic Research (Future functional materials under extreme conditions--2021B0301030005).
	\end{acknowledgments}
	
\providecommand{\noopsort}[1]{}\providecommand{\singleletter}[1]{#1}%


\begin{thebibliography}{62}%
	\makeatletter
	\providecommand \@ifxundefined [1]{%
		\@ifx{#1\undefined}
	}%
	\providecommand \@ifnum [1]{%
		\ifnum #1\expandafter \@firstoftwo
		\else \expandafter \@secondoftwo
		\fi
	}%
	\providecommand \@ifx [1]{%
		\ifx #1\expandafter \@firstoftwo
		\else \expandafter \@secondoftwo
		\fi
	}%
	\providecommand \natexlab [1]{#1}%
	\providecommand \enquote  [1]{``#1''}%
	\providecommand \bibnamefont  [1]{#1}%
	\providecommand \bibfnamefont [1]{#1}%
	\providecommand \citenamefont [1]{#1}%
	\providecommand \href@noop [0]{\@secondoftwo}%
	\providecommand \href [0]{\begingroup \@sanitize@url \@href}%
	\providecommand \@href[1]{\@@startlink{#1}\@@href}%
	\providecommand \@@href[1]{\endgroup#1\@@endlink}%
	\providecommand \@sanitize@url [0]{\catcode `\\12\catcode `\$12\catcode
		`\&12\catcode `\#12\catcode `\^12\catcode `\_12\catcode `\%12\relax}%
	\providecommand \@@startlink[1]{}%
	\providecommand \@@endlink[0]{}%
	\providecommand \url  [0]{\begingroup\@sanitize@url \@url }%
	\providecommand \@url [1]{\endgroup\@href {#1}{\urlprefix }}%
	\providecommand \urlprefix  [0]{URL }%
	\providecommand \Eprint [0]{\href }%
	\providecommand \doibase [0]{https://doi.org/}%
	\providecommand \selectlanguage [0]{\@gobble}%
	\providecommand \bibinfo  [0]{\@secondoftwo}%
	\providecommand \bibfield  [0]{\@secondoftwo}%
	\providecommand \translation [1]{[#1]}%
	\providecommand \BibitemOpen [0]{}%
	\providecommand \bibitemStop [0]{}%
	\providecommand \bibitemNoStop [0]{.\EOS\space}%
	\providecommand \EOS [0]{\spacefactor3000\relax}%
	\providecommand \BibitemShut  [1]{\csname bibitem#1\endcsname}%
	\let\auto@bib@innerbib\@empty
	%</preamble>
	\bibitem [{\citenamefont {Wolf}\ \emph {et~al.}(2001)\citenamefont {Wolf},
		\citenamefont {Awschalom}, \citenamefont {Buhrman}, \citenamefont {Daughton},
		\citenamefont {von Molnar}, \citenamefont {Roukes}, \citenamefont
		{Chtchelkanova},\ and\ \citenamefont {Treger}}]{WOS001}%
	\BibitemOpen
	\bibfield  {author} {\bibinfo {author} {\bibfnamefont {S.}~\bibnamefont
			{Wolf}}, \bibinfo {author} {\bibfnamefont {D.}~\bibnamefont {Awschalom}},
		\bibinfo {author} {\bibfnamefont {R.}~\bibnamefont {Buhrman}}, \bibinfo
		{author} {\bibfnamefont {J.}~\bibnamefont {Daughton}}, \bibinfo {author}
		{\bibfnamefont {S.}~\bibnamefont {von Molnar}}, \bibinfo {author}
		{\bibfnamefont {M.}~\bibnamefont {Roukes}}, \bibinfo {author} {\bibfnamefont
			{A.}~\bibnamefont {Chtchelkanova}},\ and\ \bibinfo {author} {\bibfnamefont
			{D.}~\bibnamefont {Treger}},\ }\bibfield  {title} {\bibinfo {title}
		{Spintronics: A spin-based electronics vision for the future},\ }\href
	{https://doi.org/10.1126/science.1065389} {\bibfield  {journal} {\bibinfo
			{journal} {SCIENCE}\ }\textbf {\bibinfo {volume} {294}},\ \bibinfo {pages}
		{1488} (\bibinfo {year} {2001})}\BibitemShut {NoStop}%
	\bibitem [{\citenamefont {Jungwirth}\ \emph {et~al.}(2016)\citenamefont
		{Jungwirth}, \citenamefont {Marti}, \citenamefont {Wadley},\ and\
		\citenamefont {Wunderlich}}]{WOS002}%
	\BibitemOpen
	\bibfield  {author} {\bibinfo {author} {\bibfnamefont {T.}~\bibnamefont
			{Jungwirth}}, \bibinfo {author} {\bibfnamefont {X.}~\bibnamefont {Marti}},
		\bibinfo {author} {\bibfnamefont {P.}~\bibnamefont {Wadley}},\ and\ \bibinfo
		{author} {\bibfnamefont {J.}~\bibnamefont {Wunderlich}},\ }\bibfield  {title}
	{\bibinfo {title} {Antiferromagnetic spintronics},\ }\href
	{https://doi.org/10.1038/NNANO.2016.18} {\bibfield  {journal} {\bibinfo
			{journal} {NATURE NANOTECHNOLOGY}\ }\textbf {\bibinfo {volume} {11}},\
		\bibinfo {pages} {231} (\bibinfo {year} {2016})}\BibitemShut {NoStop}%
	\bibitem [{\citenamefont {Smejkal}\ \emph {et~al.}(2018)\citenamefont
		{Smejkal}, \citenamefont {Mokrousov}, \citenamefont {Yan},\ and\
		\citenamefont {MacDonald}}]{WOS003}%
	\BibitemOpen
	\bibfield  {author} {\bibinfo {author} {\bibfnamefont {L.}~\bibnamefont
			{Smejkal}}, \bibinfo {author} {\bibfnamefont {Y.}~\bibnamefont {Mokrousov}},
		\bibinfo {author} {\bibfnamefont {B.}~\bibnamefont {Yan}},\ and\ \bibinfo
		{author} {\bibfnamefont {A.~H.}\ \bibnamefont {MacDonald}},\ }\bibfield
	{title} {\bibinfo {title} {Topological antiferromagnetic spintronics},\
	}\href {https://doi.org/10.1038/s41567-018-0064-5} {\bibfield  {journal}
		{\bibinfo  {journal} {NATURE PHYSICS}\ }\textbf {\bibinfo {volume} {14}},\
		\bibinfo {pages} {242} (\bibinfo {year} {2018})}\BibitemShut {NoStop}%
	\bibitem [{\citenamefont {Zutic}\ \emph {et~al.}(2004)\citenamefont {Zutic},
		\citenamefont {Fabian},\ and\ \citenamefont {Das~Sarma}}]{WOS004}%
	\BibitemOpen
	\bibfield  {author} {\bibinfo {author} {\bibfnamefont {I.}~\bibnamefont
			{Zutic}}, \bibinfo {author} {\bibfnamefont {J.}~\bibnamefont {Fabian}},\ and\
		\bibinfo {author} {\bibfnamefont {S.}~\bibnamefont {Das~Sarma}},\ }\bibfield
	{title} {\bibinfo {title} {Spintronics: Fundamentals and applications},\
	}\href {https://doi.org/10.1103/revmodphys.76.323} {\bibfield  {journal}
		{\bibinfo  {journal} {REVIEWS OF MODERN PHYSICS}\ }\textbf {\bibinfo {volume}
			{76}},\ \bibinfo {pages} {323} (\bibinfo {year} {2004})}\BibitemShut
	{NoStop}%
	\bibitem [{\citenamefont {Cherepanov}\ \emph {et~al.}(1993)\citenamefont
		{Cherepanov}, \citenamefont {Kolokolov},\ and\ \citenamefont
		{L'vov}}]{RN164}%
	\BibitemOpen
	\bibfield  {author} {\bibinfo {author} {\bibfnamefont {V.}~\bibnamefont
			{Cherepanov}}, \bibinfo {author} {\bibfnamefont {I.}~\bibnamefont
			{Kolokolov}},\ and\ \bibinfo {author} {\bibfnamefont {V.}~\bibnamefont
			{L'vov}},\ }\bibfield  {title} {\bibinfo {title} {The saga of yig: Spectra,
			thermodynamics, interaction and relaxation of magnons in a complex magnet},\
	}\href {https://doi.org/https://doi.org/10.1016/0370-1573(93)90107-O}
	{\bibfield  {journal} {\bibinfo  {journal} {Physics Reports}\ }\textbf
		{\bibinfo {volume} {229}},\ \bibinfo {pages} {81} (\bibinfo {year}
		{1993})}\BibitemShut {NoStop}%
	\bibitem [{\citenamefont {Serga}\ \emph {et~al.}(2010)\citenamefont {Serga},
		\citenamefont {Chumak},\ and\ \citenamefont {Hillebrands}}]{RN163}%
	\BibitemOpen
	\bibfield  {author} {\bibinfo {author} {\bibfnamefont {A.~A.}\ \bibnamefont
			{Serga}}, \bibinfo {author} {\bibfnamefont {A.~V.}\ \bibnamefont {Chumak}},\
		and\ \bibinfo {author} {\bibfnamefont {B.}~\bibnamefont {Hillebrands}},\
	}\bibfield  {title} {\bibinfo {title} {Yig magnonics},\ }\href
	{https://doi.org/10.1088/0022-3727/43/26/264002} {\bibfield  {journal}
		{\bibinfo  {journal} {Journal of Physics D: Applied Physics}\ }\textbf
		{\bibinfo {volume} {43}},\ \bibinfo {pages} {264002} (\bibinfo {year}
		{2010})}\BibitemShut {NoStop}%
	\bibitem [{\citenamefont {Wang}\ \emph {et~al.}(2015)\citenamefont {Wang},
		\citenamefont {Du}, \citenamefont {Hammel},\ and\ \citenamefont
		{Yang}}]{RN285}%
	\BibitemOpen
	\bibfield  {author} {\bibinfo {author} {\bibfnamefont {H.}~\bibnamefont
			{Wang}}, \bibinfo {author} {\bibfnamefont {C.}~\bibnamefont {Du}}, \bibinfo
		{author} {\bibfnamefont {P.~C.}\ \bibnamefont {Hammel}},\ and\ \bibinfo
		{author} {\bibfnamefont {F.}~\bibnamefont {Yang}},\ }\bibfield  {title}
	{\bibinfo {title} {Spin transport in antiferromagnetic insulators mediated by
			magnetic correlations},\ }\href {https://doi.org/10.1103/PhysRevB.91.220410}
	{\bibfield  {journal} {\bibinfo  {journal} {Phys. Rev. B}\ }\textbf {\bibinfo
			{volume} {91}},\ \bibinfo {pages} {220410} (\bibinfo {year}
		{2015})}\BibitemShut {NoStop}%
	\bibitem [{\citenamefont {Landau}\ and\ \citenamefont
		{Lifshitz}(1935)}]{landau1935}%
	\BibitemOpen
	\bibfield  {author} {\bibinfo {author} {\bibfnamefont {L.~D.}\ \bibnamefont
			{Landau}}\ and\ \bibinfo {author} {\bibfnamefont {E.~M.}\ \bibnamefont
			{Lifshitz}},\ }\bibfield  {title} {\bibinfo {title} {On the theory of the
			dispersion of magnetic permeability in ferromagnetic bodies},\ }\href@noop {}
	{\bibfield  {journal} {\bibinfo  {journal} {Physikalische Zeitschrift der
				Sowjetunion}\ }\textbf {\bibinfo {volume} {8}},\ \bibinfo {pages} {153}
		(\bibinfo {year} {1935})}\BibitemShut {NoStop}%
	\bibitem [{\citenamefont {Gilbert}(2004)}]{RN157}%
	\BibitemOpen
	\bibfield  {author} {\bibinfo {author} {\bibfnamefont {T.}~\bibnamefont
			{Gilbert}},\ }\bibfield  {title} {\bibinfo {title} {A phenomenological theory
			of damping in ferromagnetic materials},\ }\href
	{https://doi.org/10.1109/TMAG.2004.836740} {\bibfield  {journal} {\bibinfo
			{journal} {IEEE Transactions on Magnetics}\ }\textbf {\bibinfo {volume}
			{40}},\ \bibinfo {pages} {3443} (\bibinfo {year} {2004})}\BibitemShut
	{NoStop}%
	\bibitem [{\citenamefont {Karakurt}\ \emph {et~al.}(2007)\citenamefont
		{Karakurt}, \citenamefont {Chantrell},\ and\ \citenamefont {Nowak}}]{RN146}%
	\BibitemOpen
	\bibfield  {author} {\bibinfo {author} {\bibfnamefont {S.}~\bibnamefont
			{Karakurt}}, \bibinfo {author} {\bibfnamefont {R.}~\bibnamefont
			{Chantrell}},\ and\ \bibinfo {author} {\bibfnamefont {U.}~\bibnamefont
			{Nowak}},\ }\bibfield  {title} {\bibinfo {title} {A model of damping due to
			spin–lattice interaction},\ }\href
	{https://doi.org/https://doi.org/10.1016/j.jmmm.2007.02.118} {\bibfield
		{journal} {\bibinfo  {journal} {Journal of Magnetism and Magnetic Materials}\
		}\textbf {\bibinfo {volume} {316}},\ \bibinfo {pages} {e280} (\bibinfo {year}
		{2007})},\ \bibinfo {note} {proceedings of the Joint European Magnetic
		Symposia}\BibitemShut {NoStop}%
	\bibitem [{\citenamefont {Hickey}\ and\ \citenamefont {Moodera}(2009)}]{RN156}%
	\BibitemOpen
	\bibfield  {author} {\bibinfo {author} {\bibfnamefont {M.~C.}\ \bibnamefont
			{Hickey}}\ and\ \bibinfo {author} {\bibfnamefont {J.~S.}\ \bibnamefont
			{Moodera}},\ }\bibfield  {title} {\bibinfo {title} {Origin of intrinsic
			gilbert damping},\ }\href {https://doi.org/10.1103/PhysRevLett.102.137601}
	{\bibfield  {journal} {\bibinfo  {journal} {Phys. Rev. Lett.}\ }\textbf
		{\bibinfo {volume} {102}},\ \bibinfo {pages} {137601} (\bibinfo {year}
		{2009})}\BibitemShut {NoStop}%
	\bibitem [{\citenamefont {Widom}\ \emph {et~al.}(2010)\citenamefont {Widom},
		\citenamefont {Vittoria},\ and\ \citenamefont {Yoon}}]{RN206}%
	\BibitemOpen
	\bibfield  {author} {\bibinfo {author} {\bibfnamefont {A.}~\bibnamefont
			{Widom}}, \bibinfo {author} {\bibfnamefont {C.}~\bibnamefont {Vittoria}},\
		and\ \bibinfo {author} {\bibfnamefont {S.}~\bibnamefont {Yoon}},\ }\bibfield
	{title} {\bibinfo {title} {Gilbert ferromagnetic damping theory and the
			fluctuation-dissipation theorem},\ }\href
	{https://pubs.aip.org/aip/jap/article/108/7/073924/347820/Gilbert-ferromagnetic-damping-theory-and-the}
	{\bibfield  {journal} {\bibinfo  {journal} {Journal of Applied Physics}\
		}\textbf {\bibinfo {volume} {108}},\ \bibinfo {pages} {073924} (\bibinfo
		{year} {2010})}\BibitemShut {NoStop}%
	\bibitem [{\citenamefont {Vittoria}\ \emph {et~al.}(2010)\citenamefont
		{Vittoria}, \citenamefont {Yoon},\ and\ \citenamefont {Widom}}]{RN209}%
	\BibitemOpen
	\bibfield  {author} {\bibinfo {author} {\bibfnamefont {C.}~\bibnamefont
			{Vittoria}}, \bibinfo {author} {\bibfnamefont {S.~D.}\ \bibnamefont {Yoon}},\
		and\ \bibinfo {author} {\bibfnamefont {A.}~\bibnamefont {Widom}},\ }\bibfield
	{title} {\bibinfo {title} {Relaxation mechanism for ordered magnetic
			materials},\ }\href {https://doi.org/10.1103/PhysRevB.81.014412} {\bibfield
		{journal} {\bibinfo  {journal} {Phys. Rev. B}\ }\textbf {\bibinfo {volume}
			{81}},\ \bibinfo {pages} {014412} (\bibinfo {year} {2010})}\BibitemShut
	{NoStop}%
	\bibitem [{\citenamefont {Eriksson}\ \emph {et~al.}(2017)\citenamefont
		{Eriksson}, \citenamefont {Bergman}, \citenamefont {Bergqvist},\ and\
		\citenamefont {Hellsvik}}]{RN138}%
	\BibitemOpen
	\bibfield  {author} {\bibinfo {author} {\bibfnamefont {O.}~\bibnamefont
			{Eriksson}}, \bibinfo {author} {\bibfnamefont {A.}~\bibnamefont {Bergman}},
		\bibinfo {author} {\bibfnamefont {L.}~\bibnamefont {Bergqvist}},\ and\
		\bibinfo {author} {\bibfnamefont {J.}~\bibnamefont {Hellsvik}},\ }\href@noop
	{} {\emph {\bibinfo {title} {Atomistic spin dynamics: Foundations and
				applications}}}\ (\bibinfo  {publisher} {Oxford university press},\ \bibinfo
	{year} {2017})\BibitemShut {NoStop}%
	\bibitem [{\citenamefont {Evans}\ \emph {et~al.}(2014)\citenamefont {Evans},
		\citenamefont {Fan}, \citenamefont {Chureemart}, \citenamefont {Ostler},
		\citenamefont {Ellis},\ and\ \citenamefont {Chantrell}}]{RN345}%
	\BibitemOpen
	\bibfield  {author} {\bibinfo {author} {\bibfnamefont {R.~F.~L.}\
			\bibnamefont {Evans}}, \bibinfo {author} {\bibfnamefont {W.~J.}\ \bibnamefont
			{Fan}}, \bibinfo {author} {\bibfnamefont {P.}~\bibnamefont {Chureemart}},
		\bibinfo {author} {\bibfnamefont {T.~A.}\ \bibnamefont {Ostler}}, \bibinfo
		{author} {\bibfnamefont {M.~O.~A.}\ \bibnamefont {Ellis}},\ and\ \bibinfo
		{author} {\bibfnamefont {R.~W.}\ \bibnamefont {Chantrell}},\ }\bibfield
	{title} {\bibinfo {title} {Atomistic spin model simulations of magnetic
			nanomaterials},\ }\href {https://doi.org/10.1088/0953-8984/26/10/103202}
	{\bibfield  {journal} {\bibinfo  {journal} {Journal of Physics: Condensed
				Matter}\ }\textbf {\bibinfo {volume} {26}},\ \bibinfo {pages} {103202}
		(\bibinfo {year} {2014})}\BibitemShut {NoStop}%
	\bibitem [{\citenamefont {Kamberský}(1976)}]{RN308}%
	\BibitemOpen
	\bibfield  {author} {\bibinfo {author} {\bibfnamefont {V.}~\bibnamefont
			{Kamberský}},\ }\bibfield  {title} {\bibinfo {title} {On ferromagnetic
			resonance damping in metals},\ }\href
	{https://link.springer.com/article/10.1007/BF01587621} {\bibfield  {journal}
		{\bibinfo  {journal} {Czechoslovak Journal of Physics B}\ }\textbf {\bibinfo
			{volume} {26}},\ \bibinfo {pages} {1366} (\bibinfo {year}
		{1976})}\BibitemShut {NoStop}%
	\bibitem [{\citenamefont {Steiauf}\ and\ \citenamefont
		{F\"ahnle}(2005)}]{RN310}%
	\BibitemOpen
	\bibfield  {author} {\bibinfo {author} {\bibfnamefont {D.}~\bibnamefont
			{Steiauf}}\ and\ \bibinfo {author} {\bibfnamefont {M.}~\bibnamefont
			{F\"ahnle}},\ }\bibfield  {title} {\bibinfo {title} {Damping of spin dynamics
			in nanostructures: An ab initio study},\ }\href
	{https://doi.org/10.1103/PhysRevB.72.064450} {\bibfield  {journal} {\bibinfo
			{journal} {Phys. Rev. B}\ }\textbf {\bibinfo {volume} {72}},\ \bibinfo
		{pages} {064450} (\bibinfo {year} {2005})}\BibitemShut {NoStop}%
	\bibitem [{\citenamefont {Gilmore}\ \emph {et~al.}(2007)\citenamefont
		{Gilmore}, \citenamefont {Idzerda},\ and\ \citenamefont {Stiles}}]{RN311}%
	\BibitemOpen
	\bibfield  {author} {\bibinfo {author} {\bibfnamefont {K.}~\bibnamefont
			{Gilmore}}, \bibinfo {author} {\bibfnamefont {Y.~U.}\ \bibnamefont
			{Idzerda}},\ and\ \bibinfo {author} {\bibfnamefont {M.~D.}\ \bibnamefont
			{Stiles}},\ }\bibfield  {title} {\bibinfo {title} {Identification of the
			dominant precession-damping mechanism in fe, co, and ni by first-principles
			calculations},\ }\href {https://doi.org/10.1103/PhysRevLett.99.027204}
	{\bibfield  {journal} {\bibinfo  {journal} {Phys. Rev. Lett.}\ }\textbf
		{\bibinfo {volume} {99}},\ \bibinfo {pages} {027204} (\bibinfo {year}
		{2007})}\BibitemShut {NoStop}%
	\bibitem [{\citenamefont {Kambersk\'y}(2007)}]{RN309}%
	\BibitemOpen
	\bibfield  {author} {\bibinfo {author} {\bibfnamefont {V.}~\bibnamefont
			{Kambersk\'y}},\ }\bibfield  {title} {\bibinfo {title} {Spin-orbital gilbert
			damping in common magnetic metals},\ }\href
	{https://doi.org/10.1103/PhysRevB.76.134416} {\bibfield  {journal} {\bibinfo
			{journal} {Phys. Rev. B}\ }\textbf {\bibinfo {volume} {76}},\ \bibinfo
		{pages} {134416} (\bibinfo {year} {2007})}\BibitemShut {NoStop}%
	\bibitem [{\citenamefont {Brataas}\ \emph {et~al.}(2008)\citenamefont
		{Brataas}, \citenamefont {Tserkovnyak},\ and\ \citenamefont {Bauer}}]{RN143}%
	\BibitemOpen
	\bibfield  {author} {\bibinfo {author} {\bibfnamefont {A.}~\bibnamefont
			{Brataas}}, \bibinfo {author} {\bibfnamefont {Y.}~\bibnamefont
			{Tserkovnyak}},\ and\ \bibinfo {author} {\bibfnamefont {G.~E.~W.}\
			\bibnamefont {Bauer}},\ }\bibfield  {title} {\bibinfo {title} {Scattering
			theory of gilbert damping},\ }\href
	{https://doi.org/10.1103/PhysRevLett.101.037207} {\bibfield  {journal}
		{\bibinfo  {journal} {Phys. Rev. Lett.}\ }\textbf {\bibinfo {volume} {101}},\
		\bibinfo {pages} {037207} (\bibinfo {year} {2008})}\BibitemShut {NoStop}%
	\bibitem [{\citenamefont {Ebert}\ \emph {et~al.}(2011)\citenamefont {Ebert},
		\citenamefont {Mankovsky}, \citenamefont {Ködderitzsch},\ and\ \citenamefont
		{Kelly}}]{RN343}%
	\BibitemOpen
	\bibfield  {author} {\bibinfo {author} {\bibfnamefont {H.}~\bibnamefont
			{Ebert}}, \bibinfo {author} {\bibfnamefont {S.}~\bibnamefont {Mankovsky}},
		\bibinfo {author} {\bibfnamefont {D.}~\bibnamefont {Ködderitzsch}},\ and\
		\bibinfo {author} {\bibfnamefont {P.~J.}\ \bibnamefont {Kelly}},\ }\bibfield
	{title} {\bibinfo {title} {Ab initio calculation of the gilbert damping
			parameter via the linear response formalism},\ }\href
	{https://doi.org/10.1103/PhysRevLett.107.066603} {\bibfield  {journal}
		{\bibinfo  {journal} {Physical Review Letters}\ }\textbf {\bibinfo {volume}
			{107}},\ \bibinfo {pages} {066603} (\bibinfo {year} {2011})}\BibitemShut
	{NoStop}%
	\bibitem [{\citenamefont {Mankovsky}\ \emph {et~al.}(2013)\citenamefont
		{Mankovsky}, \citenamefont {Ködderitzsch}, \citenamefont {Woltersdorf},\
		and\ \citenamefont {Ebert}}]{RN344}%
	\BibitemOpen
	\bibfield  {author} {\bibinfo {author} {\bibfnamefont {S.}~\bibnamefont
			{Mankovsky}}, \bibinfo {author} {\bibfnamefont {D.}~\bibnamefont
			{Ködderitzsch}}, \bibinfo {author} {\bibfnamefont {G.}~\bibnamefont
			{Woltersdorf}},\ and\ \bibinfo {author} {\bibfnamefont {H.}~\bibnamefont
			{Ebert}},\ }\bibfield  {title} {\bibinfo {title} {First-principles
			calculation of the gilbert damping parameter via the linear response
			formalism with application to magnetic transition metals and alloys},\ }\href
	{https://doi.org/10.1103/PhysRevB.87.014430} {\bibfield  {journal} {\bibinfo
			{journal} {Physical Review B}\ }\textbf {\bibinfo {volume} {87}},\ \bibinfo
		{pages} {014430} (\bibinfo {year} {2013})}\BibitemShut {NoStop}%
	\bibitem [{\citenamefont {Aßmann}\ and\ \citenamefont {Nowak}(2019)}]{RN267}%
	\BibitemOpen
	\bibfield  {author} {\bibinfo {author} {\bibfnamefont {M.}~\bibnamefont
			{Aßmann}}\ and\ \bibinfo {author} {\bibfnamefont {U.}~\bibnamefont
			{Nowak}},\ }\bibfield  {title} {\bibinfo {title} {Spin-lattice relaxation
			beyond gilbert damping},\ }\href
	{https://doi.org/https://doi.org/10.1016/j.jmmm.2018.08.034} {\bibfield
		{journal} {\bibinfo  {journal} {Journal of Magnetism and Magnetic Materials}\
		}\textbf {\bibinfo {volume} {469}},\ \bibinfo {pages} {217} (\bibinfo {year}
		{2019})}\BibitemShut {NoStop}%
	\bibitem [{\citenamefont {Tranchida}\ \emph {et~al.}(2018)\citenamefont
		{Tranchida}, \citenamefont {Plimpton}, \citenamefont {Thibaudeau},\ and\
		\citenamefont {Thompson}}]{tranchida2018}%
	\BibitemOpen
	\bibfield  {author} {\bibinfo {author} {\bibfnamefont {J.}~\bibnamefont
			{Tranchida}}, \bibinfo {author} {\bibfnamefont {S.}~\bibnamefont {Plimpton}},
		\bibinfo {author} {\bibfnamefont {P.}~\bibnamefont {Thibaudeau}},\ and\
		\bibinfo {author} {\bibfnamefont {A.}~\bibnamefont {Thompson}},\ }\bibfield
	{title} {\bibinfo {title} {Massively parallel symplectic algorithm for
			coupled magnetic spin dynamics and molecular dynamics},\ }\href
	{https://doi.org/https://doi.org/10.1016/j.jcp.2018.06.042} {\bibfield
		{journal} {\bibinfo  {journal} {Journal of Computational Physics}\ }\textbf
		{\bibinfo {volume} {372}},\ \bibinfo {pages} {406} (\bibinfo {year}
		{2018})}\BibitemShut {NoStop}%
	\bibitem [{\citenamefont {Strungaru}\ \emph {et~al.}(2021)\citenamefont
		{Strungaru}, \citenamefont {Ellis}, \citenamefont {Ruta}, \citenamefont
		{Chubykalo-Fesenko}, \citenamefont {Evans},\ and\ \citenamefont
		{Chantrell}}]{RN269}%
	\BibitemOpen
	\bibfield  {author} {\bibinfo {author} {\bibfnamefont {M.}~\bibnamefont
			{Strungaru}}, \bibinfo {author} {\bibfnamefont {M.~O.~A.}\ \bibnamefont
			{Ellis}}, \bibinfo {author} {\bibfnamefont {S.}~\bibnamefont {Ruta}},
		\bibinfo {author} {\bibfnamefont {O.}~\bibnamefont {Chubykalo-Fesenko}},
		\bibinfo {author} {\bibfnamefont {R.~F.~L.}\ \bibnamefont {Evans}},\ and\
		\bibinfo {author} {\bibfnamefont {R.~W.}\ \bibnamefont {Chantrell}},\
	}\bibfield  {title} {\bibinfo {title} {Spin-lattice dynamics model with
			angular momentum transfer for canonical and microcanonical ensembles},\
	}\href {https://doi.org/10.1103/PhysRevB.103.024429} {\bibfield  {journal}
		{\bibinfo  {journal} {Phys. Rev. B}\ }\textbf {\bibinfo {volume} {103}},\
		\bibinfo {pages} {024429} (\bibinfo {year} {2021})}\BibitemShut {NoStop}%
	\bibitem [{\citenamefont {Kasuya}\ and\ \citenamefont {LeCraw}(1961)}]{RN306}%
	\BibitemOpen
	\bibfield  {author} {\bibinfo {author} {\bibfnamefont {T.}~\bibnamefont
			{Kasuya}}\ and\ \bibinfo {author} {\bibfnamefont {R.~C.}\ \bibnamefont
			{LeCraw}},\ }\bibfield  {title} {\bibinfo {title} {Relaxation mechanisms in
			ferromagnetic resonance},\ }\href {https://doi.org/10.1103/PhysRevLett.6.223}
	{\bibfield  {journal} {\bibinfo  {journal} {Phys. Rev. Lett.}\ }\textbf
		{\bibinfo {volume} {6}},\ \bibinfo {pages} {223} (\bibinfo {year}
		{1961})}\BibitemShut {NoStop}%
	\bibitem [{\citenamefont {Lou}\ \emph {et~al.}(2021)\citenamefont {Lou},
		\citenamefont {Li}, \citenamefont {Ji}, \citenamefont {Yu}, \citenamefont
		{Feng}, \citenamefont {Gong},\ and\ \citenamefont {Xiang}}]{RN40}%
	\BibitemOpen
	\bibfield  {author} {\bibinfo {author} {\bibfnamefont {F.}~\bibnamefont
			{Lou}}, \bibinfo {author} {\bibfnamefont {X.}~\bibnamefont {Li}}, \bibinfo
		{author} {\bibfnamefont {J.}~\bibnamefont {Ji}}, \bibinfo {author}
		{\bibfnamefont {H.}~\bibnamefont {Yu}}, \bibinfo {author} {\bibfnamefont
			{J.}~\bibnamefont {Feng}}, \bibinfo {author} {\bibfnamefont {X.}~\bibnamefont
			{Gong}},\ and\ \bibinfo {author} {\bibfnamefont {H.}~\bibnamefont {Xiang}},\
	}\bibfield  {title} {\bibinfo {title} {Pasp: Property analysis and simulation
			package for materials},\ }\href
	{https://pubs.aip.org/aip/jcp/article/154/11/114103/315326/PASP-Property-analysis-and-simulation-package-for}
	{\bibfield  {journal} {\bibinfo  {journal} {The Journal of Chemical Physics}\
		}\textbf {\bibinfo {volume} {154}},\ \bibinfo {pages} {114103} (\bibinfo
		{year} {2021})}\BibitemShut {NoStop}%
	\bibitem [{\citenamefont {Xiang}\ \emph {et~al.}(2011)\citenamefont {Xiang},
		\citenamefont {Kan}, \citenamefont {Wei}, \citenamefont {Whangbo},\ and\
		\citenamefont {Gong}}]{RN288}%
	\BibitemOpen
	\bibfield  {author} {\bibinfo {author} {\bibfnamefont {H.~J.}\ \bibnamefont
			{Xiang}}, \bibinfo {author} {\bibfnamefont {E.~J.}\ \bibnamefont {Kan}},
		\bibinfo {author} {\bibfnamefont {S.-H.}\ \bibnamefont {Wei}}, \bibinfo
		{author} {\bibfnamefont {M.-H.}\ \bibnamefont {Whangbo}},\ and\ \bibinfo
		{author} {\bibfnamefont {X.~G.}\ \bibnamefont {Gong}},\ }\bibfield  {title}
	{\bibinfo {title} {Predicting the spin-lattice order of frustrated systems
			from first principles},\ }\href {https://doi.org/10.1103/PhysRevB.84.224429}
	{\bibfield  {journal} {\bibinfo  {journal} {Phys. Rev. B}\ }\textbf {\bibinfo
			{volume} {84}},\ \bibinfo {pages} {224429} (\bibinfo {year}
		{2011})}\BibitemShut {NoStop}%
	\bibitem [{\citenamefont {Xiang}\ \emph {et~al.}(2013)\citenamefont {Xiang},
		\citenamefont {Lee}, \citenamefont {Koo}, \citenamefont {Gong},\ and\
		\citenamefont {Whangbo}}]{RN96}%
	\BibitemOpen
	\bibfield  {author} {\bibinfo {author} {\bibfnamefont {H.}~\bibnamefont
			{Xiang}}, \bibinfo {author} {\bibfnamefont {C.}~\bibnamefont {Lee}}, \bibinfo
		{author} {\bibfnamefont {H.-J.}\ \bibnamefont {Koo}}, \bibinfo {author}
		{\bibfnamefont {X.}~\bibnamefont {Gong}},\ and\ \bibinfo {author}
		{\bibfnamefont {M.-H.}\ \bibnamefont {Whangbo}},\ }\bibfield  {title}
	{\bibinfo {title} {Magnetic properties and energy-mapping analysis},\
	}\href@noop {} {\bibfield  {journal} {\bibinfo  {journal} {Dalton
				Transactions}\ }\textbf {\bibinfo {volume} {42}},\ \bibinfo {pages} {823}
		(\bibinfo {year} {2013})}\BibitemShut {NoStop}%
	\bibitem [{\citenamefont {Prodan}\ and\ \citenamefont {Kohn}(2005)}]{RN291}%
	\BibitemOpen
	\bibfield  {author} {\bibinfo {author} {\bibfnamefont {E.}~\bibnamefont
			{Prodan}}\ and\ \bibinfo {author} {\bibfnamefont {W.}~\bibnamefont {Kohn}},\
	}\bibfield  {title} {\bibinfo {title} {Nearsightedness of electronic
			matter},\ }\href {https://doi.org/10.1073/pnas.0505436102} {\bibfield
		{journal} {\bibinfo  {journal} {Proceedings of the National Academy of
				Sciences}\ }\textbf {\bibinfo {volume} {102}},\ \bibinfo {pages} {11635}
		(\bibinfo {year} {2005})}\BibitemShut {NoStop}%
	\bibitem [{\citenamefont {Ko}\ \emph {et~al.}(2021)\citenamefont {Ko},
		\citenamefont {Finkler}, \citenamefont {Goedecker},\ and\ \citenamefont
		{Behler}}]{RN79}%
	\BibitemOpen
	\bibfield  {author} {\bibinfo {author} {\bibfnamefont {T.~W.}\ \bibnamefont
			{Ko}}, \bibinfo {author} {\bibfnamefont {J.~A.}\ \bibnamefont {Finkler}},
		\bibinfo {author} {\bibfnamefont {S.}~\bibnamefont {Goedecker}},\ and\
		\bibinfo {author} {\bibfnamefont {J.}~\bibnamefont {Behler}},\ }\bibfield
	{title} {\bibinfo {title} {A fourth-generation high-dimensional neural
			network potential with accurate electrostatics including non-local charge
			transfer},\ }\href
	{https://doi.org/https://doi.org/10.1038/s41467-020-20427-2} {\bibfield
		{journal} {\bibinfo  {journal} {Nature communications}\ }\textbf {\bibinfo
			{volume} {12}},\ \bibinfo {pages} {1} (\bibinfo {year} {2021})}\BibitemShut
	{NoStop}%
	\bibitem [{\citenamefont {Togo}\ and\ \citenamefont {Tanaka}(2015)}]{RN296}%
	\BibitemOpen
	\bibfield  {author} {\bibinfo {author} {\bibfnamefont {A.}~\bibnamefont
			{Togo}}\ and\ \bibinfo {author} {\bibfnamefont {I.}~\bibnamefont {Tanaka}},\
	}\bibfield  {title} {\bibinfo {title} {First principles phonon calculations
			in materials science},\ }\href
	{https://doi.org/https://doi.org/10.1016/j.scriptamat.2015.07.021} {\bibfield
		{journal} {\bibinfo  {journal} {Scripta Materialia}\ }\textbf {\bibinfo
			{volume} {108}},\ \bibinfo {pages} {1} (\bibinfo {year} {2015})}\BibitemShut
	{NoStop}%
	\bibitem [{\citenamefont {Kresse}\ and\ \citenamefont {Hafner}(1993)}]{RN293}%
	\BibitemOpen
	\bibfield  {author} {\bibinfo {author} {\bibfnamefont {G.}~\bibnamefont
			{Kresse}}\ and\ \bibinfo {author} {\bibfnamefont {J.}~\bibnamefont
			{Hafner}},\ }\bibfield  {title} {\bibinfo {title} {Ab initio molecular
			dynamics for open-shell transition metals},\ }\href
	{https://doi.org/10.1103/PhysRevB.48.13115} {\bibfield  {journal} {\bibinfo
			{journal} {Phys. Rev. B}\ }\textbf {\bibinfo {volume} {48}},\ \bibinfo
		{pages} {13115} (\bibinfo {year} {1993})}\BibitemShut {NoStop}%
	\bibitem [{\citenamefont {Kresse}\ and\ \citenamefont
		{Furthm\"uller}(1996)}]{RN294}%
	\BibitemOpen
	\bibfield  {author} {\bibinfo {author} {\bibfnamefont {G.}~\bibnamefont
			{Kresse}}\ and\ \bibinfo {author} {\bibfnamefont {J.}~\bibnamefont
			{Furthm\"uller}},\ }\bibfield  {title} {\bibinfo {title} {Efficient iterative
			schemes for ab initio total-energy calculations using a plane-wave basis
			set},\ }\href {https://doi.org/10.1103/PhysRevB.54.11169} {\bibfield
		{journal} {\bibinfo  {journal} {Phys. Rev. B}\ }\textbf {\bibinfo {volume}
			{54}},\ \bibinfo {pages} {11169} (\bibinfo {year} {1996})}\BibitemShut
	{NoStop}%
	\bibitem [{\citenamefont {Kresse}\ and\ \citenamefont
		{Furthmüller}(1996)}]{RN295}%
	\BibitemOpen
	\bibfield  {author} {\bibinfo {author} {\bibfnamefont {G.}~\bibnamefont
			{Kresse}}\ and\ \bibinfo {author} {\bibfnamefont {J.}~\bibnamefont
			{Furthmüller}},\ }\bibfield  {title} {\bibinfo {title} {Efficiency of
			ab-initio total energy calculations for metals and semiconductors using a
			plane-wave basis set},\ }\href
	{https://doi.org/https://doi.org/10.1016/0927-0256(96)00008-0} {\bibfield
		{journal} {\bibinfo  {journal} {Computational Materials Science}\ }\textbf
		{\bibinfo {volume} {6}},\ \bibinfo {pages} {15} (\bibinfo {year}
		{1996})}\BibitemShut {NoStop}%
	\bibitem [{\citenamefont {Virtanen}\ \emph {et~al.}(2020)\citenamefont
		{Virtanen}, \citenamefont {Gommers}, \citenamefont {Oliphant}, \citenamefont
		{Haberland}, \citenamefont {Reddy}, \citenamefont {Cournapeau}, \citenamefont
		{Burovski}, \citenamefont {Peterson}, \citenamefont {Weckesser},\ and\
		\citenamefont {Bright}}]{RN298}%
	\BibitemOpen
	\bibfield  {author} {\bibinfo {author} {\bibfnamefont {P.}~\bibnamefont
			{Virtanen}}, \bibinfo {author} {\bibfnamefont {R.}~\bibnamefont {Gommers}},
		\bibinfo {author} {\bibfnamefont {T.~E.}\ \bibnamefont {Oliphant}}, \bibinfo
		{author} {\bibfnamefont {M.}~\bibnamefont {Haberland}}, \bibinfo {author}
		{\bibfnamefont {T.}~\bibnamefont {Reddy}}, \bibinfo {author} {\bibfnamefont
			{D.}~\bibnamefont {Cournapeau}}, \bibinfo {author} {\bibfnamefont
			{E.}~\bibnamefont {Burovski}}, \bibinfo {author} {\bibfnamefont
			{P.}~\bibnamefont {Peterson}}, \bibinfo {author} {\bibfnamefont
			{W.}~\bibnamefont {Weckesser}},\ and\ \bibinfo {author} {\bibfnamefont
			{J.}~\bibnamefont {Bright}},\ }\bibfield  {title} {\bibinfo {title} {Scipy
			1.0: fundamental algorithms for scientific computing in python},\ }\href
	{https://www.nature.com/articles/s41592%E2%80%90019%E2%80%900686%E2%80%902}
	{\bibfield  {journal} {\bibinfo  {journal} {Nature methods}\ }\textbf
		{\bibinfo {volume} {17}},\ \bibinfo {pages} {261} (\bibinfo {year}
		{2020})}\BibitemShut {NoStop}%
	\bibitem [{\citenamefont {Nurdin}\ and\ \citenamefont {Schotte}(2000)}]{RN252}%
	\BibitemOpen
	\bibfield  {author} {\bibinfo {author} {\bibfnamefont {W.~B.}\ \bibnamefont
			{Nurdin}}\ and\ \bibinfo {author} {\bibfnamefont {K.-D.}\ \bibnamefont
			{Schotte}},\ }\bibfield  {title} {\bibinfo {title} {Dynamical temperature for
			spin systems},\ }\href {https://doi.org/10.1103/PhysRevE.61.3579} {\bibfield
		{journal} {\bibinfo  {journal} {Phys. Rev. E}\ }\textbf {\bibinfo {volume}
			{61}},\ \bibinfo {pages} {3579} (\bibinfo {year} {2000})}\BibitemShut
	{NoStop}%
	\bibitem [{\citenamefont {Mentink}\ \emph {et~al.}(2010)\citenamefont
		{Mentink}, \citenamefont {Tretyakov}, \citenamefont {Fasolino}, \citenamefont
		{Katsnelson},\ and\ \citenamefont {Rasing}}]{RN127}%
	\BibitemOpen
	\bibfield  {author} {\bibinfo {author} {\bibfnamefont {J.~H.}\ \bibnamefont
			{Mentink}}, \bibinfo {author} {\bibfnamefont {M.~V.}\ \bibnamefont
			{Tretyakov}}, \bibinfo {author} {\bibfnamefont {A.}~\bibnamefont {Fasolino}},
		\bibinfo {author} {\bibfnamefont {M.~I.}\ \bibnamefont {Katsnelson}},\ and\
		\bibinfo {author} {\bibfnamefont {T.}~\bibnamefont {Rasing}},\ }\bibfield
	{title} {\bibinfo {title} {Stable and fast semi-implicit integration of the
			stochastic landau–lifshitz equation},\ }\href
	{https://doi.org/10.1088/0953-8984/22/17/176001} {\bibfield  {journal}
		{\bibinfo  {journal} {Journal of Physics: Condensed Matter}\ }\textbf
		{\bibinfo {volume} {22}},\ \bibinfo {pages} {176001} (\bibinfo {year}
		{2010})}\BibitemShut {NoStop}%
	\bibitem [{\citenamefont {Grønbech-Jensen}\ and\ \citenamefont
		{Farago}(2013)}]{RN136}%
	\BibitemOpen
	\bibfield  {author} {\bibinfo {author} {\bibfnamefont {N.}~\bibnamefont
			{Grønbech-Jensen}}\ and\ \bibinfo {author} {\bibfnamefont {O.}~\bibnamefont
			{Farago}},\ }\bibfield  {title} {\bibinfo {title} {A simple and effective
			verlet-type algorithm for simulating langevin dynamics},\ }\href
	{https://www.tandfonline.com/doi/full/10.1080/00268976.2012.760055}
	{\bibfield  {journal} {\bibinfo  {journal} {Molecular Physics}\ }\textbf
		{\bibinfo {volume} {111}},\ \bibinfo {pages} {983} (\bibinfo {year}
		{2013})}\BibitemShut {NoStop}%
	\bibitem [{\citenamefont {Wang}\ \emph {et~al.}(2012)\citenamefont {Wang},
		\citenamefont {Weerasinghe},\ and\ \citenamefont {Bellaiche}}]{RN133}%
	\BibitemOpen
	\bibfield  {author} {\bibinfo {author} {\bibfnamefont {D.}~\bibnamefont
			{Wang}}, \bibinfo {author} {\bibfnamefont {J.}~\bibnamefont {Weerasinghe}},\
		and\ \bibinfo {author} {\bibfnamefont {L.}~\bibnamefont {Bellaiche}},\
	}\bibfield  {title} {\bibinfo {title} {Atomistic molecular dynamic
			simulations of multiferroics},\ }\href
	{https://doi.org/10.1103/PhysRevLett.109.067203} {\bibfield  {journal}
		{\bibinfo  {journal} {Phys. Rev. Lett.}\ }\textbf {\bibinfo {volume} {109}},\
		\bibinfo {pages} {067203} (\bibinfo {year} {2012})}\BibitemShut {NoStop}%
	\bibitem [{\citenamefont {Sinha}\ and\ \citenamefont
		{Upadhyaya}(1962)}]{RN154}%
	\BibitemOpen
	\bibfield  {author} {\bibinfo {author} {\bibfnamefont {K.~P.}\ \bibnamefont
			{Sinha}}\ and\ \bibinfo {author} {\bibfnamefont {U.~N.}\ \bibnamefont
			{Upadhyaya}},\ }\bibfield  {title} {\bibinfo {title} {Phonon-magnon
			interaction in magnetic crystals},\ }\href
	{https://doi.org/10.1103/PhysRev.127.432} {\bibfield  {journal} {\bibinfo
			{journal} {Phys. Rev.}\ }\textbf {\bibinfo {volume} {127}},\ \bibinfo {pages}
		{432} (\bibinfo {year} {1962})}\BibitemShut {NoStop}%
	\bibitem [{\citenamefont {Upadhyaya}\ and\ \citenamefont
		{Sinha}(1963)}]{RN145}%
	\BibitemOpen
	\bibfield  {author} {\bibinfo {author} {\bibfnamefont {U.~N.}\ \bibnamefont
			{Upadhyaya}}\ and\ \bibinfo {author} {\bibfnamefont {K.~P.}\ \bibnamefont
			{Sinha}},\ }\bibfield  {title} {\bibinfo {title} {Phonon-magnon interaction
			in magnetic crystals. ii. antiferromagnetic systems},\ }\href
	{https://doi.org/10.1103/PhysRev.130.939} {\bibfield  {journal} {\bibinfo
			{journal} {Phys. Rev.}\ }\textbf {\bibinfo {volume} {130}},\ \bibinfo {pages}
		{939} (\bibinfo {year} {1963})}\BibitemShut {NoStop}%
	\bibitem [{\citenamefont {Campbell}\ \emph {et~al.}(2020)\citenamefont
		{Campbell}, \citenamefont {Xu}, \citenamefont {Bayaraa},\ and\ \citenamefont
		{Bellaiche}}]{RN168}%
	\BibitemOpen
	\bibfield  {author} {\bibinfo {author} {\bibfnamefont {D.}~\bibnamefont
			{Campbell}}, \bibinfo {author} {\bibfnamefont {C.}~\bibnamefont {Xu}},
		\bibinfo {author} {\bibfnamefont {T.}~\bibnamefont {Bayaraa}},\ and\ \bibinfo
		{author} {\bibfnamefont {L.}~\bibnamefont {Bellaiche}},\ }\bibfield  {title}
	{\bibinfo {title} {Finite-temperature properties of rare-earth iron garnets
			in a magnetic field},\ }\href {https://doi.org/10.1103/PhysRevB.102.144406}
	{\bibfield  {journal} {\bibinfo  {journal} {Phys. Rev. B}\ }\textbf {\bibinfo
			{volume} {102}},\ \bibinfo {pages} {144406} (\bibinfo {year}
		{2020})}\BibitemShut {NoStop}%
	\bibitem [{\citenamefont {Bl\"ochl}(1994)}]{RN301}%
	\BibitemOpen
	\bibfield  {author} {\bibinfo {author} {\bibfnamefont {P.~E.}\ \bibnamefont
			{Bl\"ochl}},\ }\bibfield  {title} {\bibinfo {title} {Projector augmented-wave
			method},\ }\href {https://doi.org/10.1103/PhysRevB.50.17953} {\bibfield
		{journal} {\bibinfo  {journal} {Phys. Rev. B}\ }\textbf {\bibinfo {volume}
			{50}},\ \bibinfo {pages} {17953} (\bibinfo {year} {1994})}\BibitemShut
	{NoStop}%
	\bibitem [{\citenamefont {Perdew}\ \emph {et~al.}(2008)\citenamefont {Perdew},
		\citenamefont {Ruzsinszky}, \citenamefont {Csonka}, \citenamefont {Vydrov},
		\citenamefont {Scuseria}, \citenamefont {Constantin}, \citenamefont {Zhou},\
		and\ \citenamefont {Burke}}]{RN302}%
	\BibitemOpen
	\bibfield  {author} {\bibinfo {author} {\bibfnamefont {J.~P.}\ \bibnamefont
			{Perdew}}, \bibinfo {author} {\bibfnamefont {A.}~\bibnamefont {Ruzsinszky}},
		\bibinfo {author} {\bibfnamefont {G.~I.}\ \bibnamefont {Csonka}}, \bibinfo
		{author} {\bibfnamefont {O.~A.}\ \bibnamefont {Vydrov}}, \bibinfo {author}
		{\bibfnamefont {G.~E.}\ \bibnamefont {Scuseria}}, \bibinfo {author}
		{\bibfnamefont {L.~A.}\ \bibnamefont {Constantin}}, \bibinfo {author}
		{\bibfnamefont {X.}~\bibnamefont {Zhou}},\ and\ \bibinfo {author}
		{\bibfnamefont {K.}~\bibnamefont {Burke}},\ }\bibfield  {title} {\bibinfo
		{title} {Restoring the density-gradient expansion for exchange in solids and
			surfaces},\ }\href {https://doi.org/10.1103/PhysRevLett.100.136406}
	{\bibfield  {journal} {\bibinfo  {journal} {Phys. Rev. Lett.}\ }\textbf
		{\bibinfo {volume} {100}},\ \bibinfo {pages} {136406} (\bibinfo {year}
		{2008})}\BibitemShut {NoStop}%
	\bibitem [{\citenamefont {Dudarev}\ \emph {et~al.}(1998)\citenamefont
		{Dudarev}, \citenamefont {Botton}, \citenamefont {Savrasov}, \citenamefont
		{Humphreys},\ and\ \citenamefont {Sutton}}]{RN303}%
	\BibitemOpen
	\bibfield  {author} {\bibinfo {author} {\bibfnamefont {S.~L.}\ \bibnamefont
			{Dudarev}}, \bibinfo {author} {\bibfnamefont {G.~A.}\ \bibnamefont {Botton}},
		\bibinfo {author} {\bibfnamefont {S.~Y.}\ \bibnamefont {Savrasov}}, \bibinfo
		{author} {\bibfnamefont {C.~J.}\ \bibnamefont {Humphreys}},\ and\ \bibinfo
		{author} {\bibfnamefont {A.~P.}\ \bibnamefont {Sutton}},\ }\bibfield  {title}
	{\bibinfo {title} {Electron-energy-loss spectra and the structural stability
			of nickel oxide: An lsda+u study},\ }\href
	{https://doi.org/10.1103/PhysRevB.57.1505} {\bibfield  {journal} {\bibinfo
			{journal} {Phys. Rev. B}\ }\textbf {\bibinfo {volume} {57}},\ \bibinfo
		{pages} {1505} (\bibinfo {year} {1998})}\BibitemShut {NoStop}%
	\bibitem [{\citenamefont {Sun}\ \emph {et~al.}(2012)\citenamefont {Sun},
		\citenamefont {Song}, \citenamefont {Chang}, \citenamefont {Kabatek},
		\citenamefont {Jantz}, \citenamefont {Schneider}, \citenamefont {Wu},
		\citenamefont {Schultheiss},\ and\ \citenamefont {Hoffmann}}]{RN202}%
	\BibitemOpen
	\bibfield  {author} {\bibinfo {author} {\bibfnamefont {Y.}~\bibnamefont
			{Sun}}, \bibinfo {author} {\bibfnamefont {Y.-Y.}\ \bibnamefont {Song}},
		\bibinfo {author} {\bibfnamefont {H.}~\bibnamefont {Chang}}, \bibinfo
		{author} {\bibfnamefont {M.}~\bibnamefont {Kabatek}}, \bibinfo {author}
		{\bibfnamefont {M.}~\bibnamefont {Jantz}}, \bibinfo {author} {\bibfnamefont
			{W.}~\bibnamefont {Schneider}}, \bibinfo {author} {\bibfnamefont
			{M.}~\bibnamefont {Wu}}, \bibinfo {author} {\bibfnamefont {H.}~\bibnamefont
			{Schultheiss}},\ and\ \bibinfo {author} {\bibfnamefont {A.}~\bibnamefont
			{Hoffmann}},\ }\bibfield  {title} {\bibinfo {title} {Growth and ferromagnetic
			resonance properties of nanometer-thick yttrium iron garnet films},\ }\href
	{https://pubs.aip.org/aip/apl/article/101/15/152405/150886/Growth-and-ferromagnetic-resonance-properties-of}
	{\bibfield  {journal} {\bibinfo  {journal} {Applied Physics Letters}\
		}\textbf {\bibinfo {volume} {101}},\ \bibinfo {pages} {152405} (\bibinfo
		{year} {2012})}\BibitemShut {NoStop}%
	\bibitem [{\citenamefont {Onbasli}\ \emph {et~al.}(2014)\citenamefont
		{Onbasli}, \citenamefont {Kehlberger}, \citenamefont {Kim}, \citenamefont
		{Jakob}, \citenamefont {Kläui}, \citenamefont {Chumak}, \citenamefont
		{Hillebrands},\ and\ \citenamefont {Ross}}]{RN205}%
	\BibitemOpen
	\bibfield  {author} {\bibinfo {author} {\bibfnamefont {M.}~\bibnamefont
			{Onbasli}}, \bibinfo {author} {\bibfnamefont {A.}~\bibnamefont {Kehlberger}},
		\bibinfo {author} {\bibfnamefont {D.~H.}\ \bibnamefont {Kim}}, \bibinfo
		{author} {\bibfnamefont {G.}~\bibnamefont {Jakob}}, \bibinfo {author}
		{\bibfnamefont {M.}~\bibnamefont {Kläui}}, \bibinfo {author} {\bibfnamefont
			{A.~V.}\ \bibnamefont {Chumak}}, \bibinfo {author} {\bibfnamefont
			{B.}~\bibnamefont {Hillebrands}},\ and\ \bibinfo {author} {\bibfnamefont
			{C.~A.}\ \bibnamefont {Ross}},\ }\bibfield  {title} {\bibinfo {title} {Pulsed
			laser deposition of epitaxial yttrium iron garnet films with low gilbert
			damping and bulk-like magnetization},\ }\href
	{https://pubs.aip.org/aip/apm/article/2/10/106102/119970/Pulsed-laser-deposition-of-epitaxial-yttrium-iron}
	{\bibfield  {journal} {\bibinfo  {journal} {APL Materials}\ }\textbf
		{\bibinfo {volume} {2}},\ \bibinfo {pages} {106102} (\bibinfo {year}
		{2014})}\BibitemShut {NoStop}%
	\bibitem [{\citenamefont {Dubs}\ \emph {et~al.}(2017)\citenamefont {Dubs},
		\citenamefont {Surzhenko}, \citenamefont {Linke}, \citenamefont {Danilewsky},
		\citenamefont {Brückner},\ and\ \citenamefont {Dellith}}]{RN204}%
	\BibitemOpen
	\bibfield  {author} {\bibinfo {author} {\bibfnamefont {C.}~\bibnamefont
			{Dubs}}, \bibinfo {author} {\bibfnamefont {O.}~\bibnamefont {Surzhenko}},
		\bibinfo {author} {\bibfnamefont {R.}~\bibnamefont {Linke}}, \bibinfo
		{author} {\bibfnamefont {A.}~\bibnamefont {Danilewsky}}, \bibinfo {author}
		{\bibfnamefont {U.}~\bibnamefont {Brückner}},\ and\ \bibinfo {author}
		{\bibfnamefont {J.}~\bibnamefont {Dellith}},\ }\bibfield  {title} {\bibinfo
		{title} {Sub-micrometer yttrium iron garnet lpe films with low ferromagnetic
			resonance losses},\ }\href
	{https://iopscience.iop.org/article/10.1088/1361-6463/aa6b1c/meta} {\bibfield
		{journal} {\bibinfo  {journal} {Journal of Physics D: Applied Physics}\
		}\textbf {\bibinfo {volume} {50}},\ \bibinfo {pages} {204005} (\bibinfo
		{year} {2017})}\BibitemShut {NoStop}%
	\bibitem [{\citenamefont {Röschmann}\ and\ \citenamefont
		{Tolksdorf}(1983)}]{roschmann1983}%
	\BibitemOpen
	\bibfield  {author} {\bibinfo {author} {\bibfnamefont {P.}~\bibnamefont
			{Röschmann}}\ and\ \bibinfo {author} {\bibfnamefont {W.}~\bibnamefont
			{Tolksdorf}},\ }\bibfield  {title} {\bibinfo {title} {Epitaxial growth and
			annealing control of fmr properties of thick homogeneous ga substituted
			yttrium iron garnet films},\ }\href
	{https://doi.org/https://doi.org/10.1016/0025-5408(83)90137-X} {\bibfield
		{journal} {\bibinfo  {journal} {Materials Research Bulletin}\ }\textbf
		{\bibinfo {volume} {18}},\ \bibinfo {pages} {449} (\bibinfo {year}
		{1983})}\BibitemShut {NoStop}%
	\bibitem [{\citenamefont {Goodenough}\ \emph {et~al.}(1970)\citenamefont
		{Goodenough}, \citenamefont {Gr{\"a}per}, \citenamefont {Holtzberg},
		\citenamefont {Huber}, \citenamefont {Lefever}, \citenamefont {Longo},
		\citenamefont {McGuire},\ and\ \citenamefont {Methfessel}}]{goodenough1970}%
	\BibitemOpen
	\bibfield  {author} {\bibinfo {author} {\bibfnamefont {J.}~\bibnamefont
			{Goodenough}}, \bibinfo {author} {\bibfnamefont {W.}~\bibnamefont
			{Gr{\"a}per}}, \bibinfo {author} {\bibfnamefont {F.}~\bibnamefont
			{Holtzberg}}, \bibinfo {author} {\bibfnamefont {D.}~\bibnamefont {Huber}},
		\bibinfo {author} {\bibfnamefont {R.}~\bibnamefont {Lefever}}, \bibinfo
		{author} {\bibfnamefont {J.}~\bibnamefont {Longo}}, \bibinfo {author}
		{\bibfnamefont {T.}~\bibnamefont {McGuire}},\ and\ \bibinfo {author}
		{\bibfnamefont {S.}~\bibnamefont {Methfessel}},\ }\href@noop {} {\emph
		{\bibinfo {title} {Magnetic and other properties of oxides and related
				compounds}}}\ (\bibinfo  {publisher} {Springer},\ \bibinfo {year}
	{1970})\BibitemShut {NoStop}%
	\bibitem [{\citenamefont {Huang}\ and\ \citenamefont {Cheng}(2013)}]{RN232}%
	\BibitemOpen
	\bibfield  {author} {\bibinfo {author} {\bibfnamefont {J.-R.}\ \bibnamefont
			{Huang}}\ and\ \bibinfo {author} {\bibfnamefont {C.}~\bibnamefont {Cheng}},\
	}\bibfield  {title} {\bibinfo {title} {Cation and magnetic orders in mnfe2o4
			from density functional calculations},\ }\href
	{https://pubs.aip.org/aip/jap/article/113/3/033912/346150/Cation-and-magnetic-orders-in-MnFe2O4-from-density}
	{\bibfield  {journal} {\bibinfo  {journal} {Journal of Applied Physics}\
		}\textbf {\bibinfo {volume} {113}},\ \bibinfo {pages} {033912} (\bibinfo
		{year} {2013})}\BibitemShut {NoStop}%
	\bibitem [{\citenamefont {Flores}\ \emph {et~al.}(1999)\citenamefont {Flores},
		\citenamefont {Raposo}, \citenamefont {Torres},\ and\ \citenamefont
		{I\~niguez}}]{RN238}%
	\BibitemOpen
	\bibfield  {author} {\bibinfo {author} {\bibfnamefont {A.~G.}\ \bibnamefont
			{Flores}}, \bibinfo {author} {\bibfnamefont {V.}~\bibnamefont {Raposo}},
		\bibinfo {author} {\bibfnamefont {L.}~\bibnamefont {Torres}},\ and\ \bibinfo
		{author} {\bibfnamefont {J.}~\bibnamefont {I\~niguez}},\ }\bibfield  {title}
	{\bibinfo {title} {Two-magnon processes and ferrimagnetic linewidth
			calculation in manganese ferrite},\ }\href
	{https://doi.org/10.1103/PhysRevB.59.9447} {\bibfield  {journal} {\bibinfo
			{journal} {Phys. Rev. B}\ }\textbf {\bibinfo {volume} {59}},\ \bibinfo
		{pages} {9447} (\bibinfo {year} {1999})}\BibitemShut {NoStop}%
	\bibitem [{\citenamefont {Reddy}\ \emph {et~al.}(1987)\citenamefont {Reddy},
		\citenamefont {Pratap},\ and\ \citenamefont {Rao}}]{RN235}%
	\BibitemOpen
	\bibfield  {author} {\bibinfo {author} {\bibfnamefont {P.~V.}\ \bibnamefont
			{Reddy}}, \bibinfo {author} {\bibfnamefont {K.}~\bibnamefont {Pratap}},\ and\
		\bibinfo {author} {\bibfnamefont {T.~S.}\ \bibnamefont {Rao}},\ }\bibfield
	{title} {\bibinfo {title} {Electrical conductivity of some mixed ferrites at
			curie point},\ }\href
	{https://onlinelibrary.wiley.com/doi/abs/10.1002/crat.2170220718} {\bibfield
		{journal} {\bibinfo  {journal} {Crystal Research and Technology}\ }\textbf
		{\bibinfo {volume} {22}},\ \bibinfo {pages} {977} (\bibinfo {year}
		{1987})}\BibitemShut {NoStop}%
	\bibitem [{\citenamefont {Hedrich}\ \emph {et~al.}(2021)\citenamefont
		{Hedrich}, \citenamefont {Wagner}, \citenamefont {Pylypovskyi}, \citenamefont
		{Shields}, \citenamefont {Kosub}, \citenamefont {Sheka}, \citenamefont
		{Makarov},\ and\ \citenamefont {Maletinsky}}]{RN225}%
	\BibitemOpen
	\bibfield  {author} {\bibinfo {author} {\bibfnamefont {N.}~\bibnamefont
			{Hedrich}}, \bibinfo {author} {\bibfnamefont {K.}~\bibnamefont {Wagner}},
		\bibinfo {author} {\bibfnamefont {O.~V.}\ \bibnamefont {Pylypovskyi}},
		\bibinfo {author} {\bibfnamefont {B.~J.}\ \bibnamefont {Shields}}, \bibinfo
		{author} {\bibfnamefont {T.}~\bibnamefont {Kosub}}, \bibinfo {author}
		{\bibfnamefont {D.~D.}\ \bibnamefont {Sheka}}, \bibinfo {author}
		{\bibfnamefont {D.}~\bibnamefont {Makarov}},\ and\ \bibinfo {author}
		{\bibfnamefont {P.}~\bibnamefont {Maletinsky}},\ }\bibfield  {title}
	{\bibinfo {title} {Nanoscale mechanics of antiferromagnetic domain walls},\
	}\href {https://www.nature.com/articles/s41567-020-01157-0} {\bibfield
		{journal} {\bibinfo  {journal} {Nature Physics}\ }\textbf {\bibinfo {volume}
			{17}},\ \bibinfo {pages} {574} (\bibinfo {year} {2021})}\BibitemShut
	{NoStop}%
	\bibitem [{\citenamefont {Xiao}\ \emph {et~al.}(2022)\citenamefont {Xiao},
		\citenamefont {Zhang}, \citenamefont {Wang},\ and\ \citenamefont
		{Cheng}}]{RN257}%
	\BibitemOpen
	\bibfield  {author} {\bibinfo {author} {\bibfnamefont {W.-Z.}\ \bibnamefont
			{Xiao}}, \bibinfo {author} {\bibfnamefont {Y.-W.}\ \bibnamefont {Zhang}},
		\bibinfo {author} {\bibfnamefont {L.-L.}\ \bibnamefont {Wang}},\ and\
		\bibinfo {author} {\bibfnamefont {C.-P.}\ \bibnamefont {Cheng}},\ }\bibfield
	{title} {\bibinfo {title} {Newtype two-dimensional cr2o3 monolayer with
			half-metallicity, high curie temperature, and magnetic anisotropy},\ }\href
	{https://doi.org/https://doi.org/10.1016/j.jmmm.2021.168657} {\bibfield
		{journal} {\bibinfo  {journal} {Journal of Magnetism and Magnetic Materials}\
		}\textbf {\bibinfo {volume} {543}},\ \bibinfo {pages} {168657} (\bibinfo
		{year} {2022})}\BibitemShut {NoStop}%
	\bibitem [{\citenamefont {Makushko}\ \emph {et~al.}(2022)\citenamefont
		{Makushko}, \citenamefont {Kosub}, \citenamefont {Pylypovskyi}, \citenamefont
		{Hedrich}, \citenamefont {Li}, \citenamefont {Pashkin}, \citenamefont
		{Avdoshenko}, \citenamefont {H{\"u}bner}, \citenamefont {Ganss},
		\citenamefont {Wolf} \emph {et~al.}}]{makushko2022}%
	\BibitemOpen
	\bibfield  {author} {\bibinfo {author} {\bibfnamefont {P.}~\bibnamefont
			{Makushko}}, \bibinfo {author} {\bibfnamefont {T.}~\bibnamefont {Kosub}},
		\bibinfo {author} {\bibfnamefont {O.~V.}\ \bibnamefont {Pylypovskyi}},
		\bibinfo {author} {\bibfnamefont {N.}~\bibnamefont {Hedrich}}, \bibinfo
		{author} {\bibfnamefont {J.}~\bibnamefont {Li}}, \bibinfo {author}
		{\bibfnamefont {A.}~\bibnamefont {Pashkin}}, \bibinfo {author} {\bibfnamefont
			{S.}~\bibnamefont {Avdoshenko}}, \bibinfo {author} {\bibfnamefont
			{R.}~\bibnamefont {H{\"u}bner}}, \bibinfo {author} {\bibfnamefont
			{F.}~\bibnamefont {Ganss}}, \bibinfo {author} {\bibfnamefont
			{D.}~\bibnamefont {Wolf}}, \emph {et~al.},\ }\bibfield  {title} {\bibinfo
		{title} {Flexomagnetism and vertically graded n{\'e}el temperature of
			antiferromagnetic cr2o3 thin films},\ }\href
	{https://www.nature.com/articles/s41467-022-34233-5} {\bibfield  {journal}
		{\bibinfo  {journal} {Nature Communications}\ }\textbf {\bibinfo {volume}
			{13}},\ \bibinfo {pages} {6745} (\bibinfo {year} {2022})}\BibitemShut
	{NoStop}%
	\bibitem [{\citenamefont {Shi}\ \emph {et~al.}(2009)\citenamefont {Shi},
		\citenamefont {Wysocki},\ and\ \citenamefont {Belashchenko}}]{RN226}%
	\BibitemOpen
	\bibfield  {author} {\bibinfo {author} {\bibfnamefont {S.}~\bibnamefont
			{Shi}}, \bibinfo {author} {\bibfnamefont {A.~L.}\ \bibnamefont {Wysocki}},\
		and\ \bibinfo {author} {\bibfnamefont {K.~D.}\ \bibnamefont {Belashchenko}},\
	}\bibfield  {title} {\bibinfo {title} {Magnetism of chromia from
			first-principles calculations},\ }\href
	{https://doi.org/10.1103/PhysRevB.79.104404} {\bibfield  {journal} {\bibinfo
			{journal} {Phys. Rev. B}\ }\textbf {\bibinfo {volume} {79}},\ \bibinfo
		{pages} {104404} (\bibinfo {year} {2009})}\BibitemShut {NoStop}%
	\bibitem [{\citenamefont {Liechtenstein}\ \emph {et~al.}(1995)\citenamefont
		{Liechtenstein}, \citenamefont {Anisimov},\ and\ \citenamefont
		{Zaanen}}]{RN304}%
	\BibitemOpen
	\bibfield  {author} {\bibinfo {author} {\bibfnamefont {A.~I.}\ \bibnamefont
			{Liechtenstein}}, \bibinfo {author} {\bibfnamefont {V.~I.}\ \bibnamefont
			{Anisimov}},\ and\ \bibinfo {author} {\bibfnamefont {J.}~\bibnamefont
			{Zaanen}},\ }\bibfield  {title} {\bibinfo {title} {Density-functional theory
			and strong interactions: Orbital ordering in mott-hubbard insulators},\
	}\href {https://doi.org/10.1103/PhysRevB.52.R5467} {\bibfield  {journal}
		{\bibinfo  {journal} {Phys. Rev. B}\ }\textbf {\bibinfo {volume} {52}},\
		\bibinfo {pages} {R5467} (\bibinfo {year} {1995})}\BibitemShut {NoStop}%
	\bibitem [{\citenamefont {Brown}(1969)}]{brown1969magneto}%
	\BibitemOpen
	\bibfield  {author} {\bibinfo {author} {\bibfnamefont {C.~A.}\ \bibnamefont
			{Brown}},\ }\emph {\bibinfo {title} {Magneto-electric domains in single
			crystal chromium oxide}},\ \href@noop {} {\bibinfo {type} {{Ph.D.} thesis}},\
	\bibinfo  {school} {Imperial College London} (\bibinfo {year}
	{1969})\BibitemShut {NoStop}%
	\bibitem [{\citenamefont {Belashchenko}\ \emph {et~al.}(2016)\citenamefont
		{Belashchenko}, \citenamefont {Tchernyshyov}, \citenamefont {Kovalev},\ and\
		\citenamefont {Tretiakov}}]{RN222}%
	\BibitemOpen
	\bibfield  {author} {\bibinfo {author} {\bibfnamefont {K.~D.}\ \bibnamefont
			{Belashchenko}}, \bibinfo {author} {\bibfnamefont {O.}~\bibnamefont
			{Tchernyshyov}}, \bibinfo {author} {\bibfnamefont {A.~A.}\ \bibnamefont
			{Kovalev}},\ and\ \bibinfo {author} {\bibfnamefont {O.~A.}\ \bibnamefont
			{Tretiakov}},\ }\bibfield  {title} {\bibinfo {title} {Magnetoelectric domain
			wall dynamics and its implications for magnetoelectric memory},\ }\href
	{https://pubs.aip.org/aip/apl/article/108/13/132403/29790/Magnetoelectric-domain-wall-dynamics-and-its}
	{\bibfield  {journal} {\bibinfo  {journal} {Applied Physics Letters}\
		}\textbf {\bibinfo {volume} {108}},\ \bibinfo {pages} {132403} (\bibinfo
		{year} {2016})}\BibitemShut {NoStop}%
	\bibitem [{\citenamefont {Li}\ \emph {et~al.}(2021)\citenamefont {Li},
		\citenamefont {Feng}, \citenamefont {Wang}, \citenamefont {Kan},\ and\
		\citenamefont {Xiang}}]{WOS005}%
	\BibitemOpen
	\bibfield  {author} {\bibinfo {author} {\bibfnamefont {J.}~\bibnamefont
			{Li}}, \bibinfo {author} {\bibfnamefont {J.}~\bibnamefont {Feng}}, \bibinfo
		{author} {\bibfnamefont {P.}~\bibnamefont {Wang}}, \bibinfo {author}
		{\bibfnamefont {E.}~\bibnamefont {Kan}},\ and\ \bibinfo {author}
		{\bibfnamefont {H.}~\bibnamefont {Xiang}},\ }\bibfield  {title} {\bibinfo
		{title} {Nature of spin-lattice coupling in two-dimensional cri3 and
			crgete3},\ }\bibfield  {journal} {\bibinfo  {journal} {SCIENCE CHINA-PHYSICS
			MECHANICS \& ASTRONOMY}\ }\textbf {\bibinfo {volume} {64}},\ \href
	{https://doi.org/10.1007/s11433-021-1717-9} {10.1007/s11433-021-1717-9}
	(\bibinfo {year} {2021})\BibitemShut {NoStop}%
\end{thebibliography}
\end{document}